\documentclass[traditabstract]{aa}
\usepackage{graphicx}
\usepackage{amsmath}
\usepackage{cancel}
\usepackage{amssymb}

\def\deg{^{\circ}}
\newcommand{\ve}[1]{{\rm\bf {#1}}}
\def\df{\textrm{d}}
\def\ex{{\bf e_x}}
\def\ez{{\bf e_z}}
\def\ey{{\bf e_y}}
\def\expr{{\bf e_x^{'}}}
\def\ezpr{{\bf e_z^{'}}}
\def\eypr{{\bf e_y^{'}}}

\begin{document}

\title{Inferring the magnetic field vector in the quiet Sun}
\subtitle{III. Disk variation of the Stokes profiles and isotropism
of the magnetic field}

\author{J.M.~Borrero\inst{1} \and P. Kobel\inst{2}}
\institute{Kiepenheuer-Institut f\"ur Sonnenphysik, Sch\"oneckstr. 6, D-79110, Freiburg, Germany. \email{borrero@kis.uni-freiburg.de}
 \and \'Ecole Polytechnique F\'ed\'erale de Lausanne, Laboratory for Hydraulic Machines, Avenue de Cour 33 Bis, 1007 Lausanne, Switzerland \email{philippe.kobel@epfl.ch}}
\date{Received / Accepted}

\begin{abstract}
{Recent investigations of the magnetic field vector properties in the solar internetwork have provided diverging results.
While some works found that the internetwork is mostly pervaded by horizontal magnetic fields, other works argued in favor of an isotropic
distribution of the magnetic field vector. Motivated by these seemingly contradictory results and by the fact that most of these works have 
employed spectropolarimetric data at disk center only, we have revisited this problem employing high-quality data (noise level $\sigma \approx 3\times
10^{-4}$ in units of the quiet-Sun intensity) at different latitudes recorded with the Hinode/SP instrument. Instead of applying traditional 
inversion codes of the radiative transfer equation to retrieve the magnetic field vector at each spatial point on the solar surface and studying 
the resulting distribution of the magnetic field vector, we surmised a theoretical distribution function of the magnetic field vector 
and used it to obtain the theoretical histograms of the Stokes profiles. These histograms were then compared to the observed ones. Any 
mismatch between them was ascribed to the theoretical distribution of the magnetic field vector, which was subsequently modified to produce 
a better fit to the observed histograms. With this method we find that Stokes profiles with signals above $2\times 10^{-3}$ (in units of 
the continuum intensity) cannot be explained by an isotropic distribution of the magnetic field vector. We also find that the differences 
between the histograms of the Stokes profiles observed at different latitudes cannot be explained in terms of line-of-sight effects. 
However, they can be explained by a distribution of the magnetic field vector that inherently varies with latitude. We note that these 
results are based on a series of assumptions that, although briefly discussed in this paper, need to be considered in more detail in the 
future.}
\end{abstract}

\maketitle

\section{Introduction}
\label{section:intro}

In recent years several attempts have been made to investigate the magnetic field vector distribution in the solar 
internetwork. Initially, these works studied the magnetic field strength is in these regions. Some favored
magnetic fields of about a few hundred Gauss or less (Asensio Ramos et al. 2007; L\'opez Ariste et al. 2007; Orozco 
Su\'arez et al. 2007a, Orozco Su\'arez \& Bellot Rubio 2012) while others found magnetic fields in the kilo-Gauss 
range (Dom{\'\i}nguez Cerde\~na et al. 2003, 2006; S\'anchez Almeida 2005). These studies were carried out mostly with 
low spatial resolution data (1"). Whenever the spatial resolution increased to better than 1 arcsec, this decreased
the signal-to-noise ratio. With the Hinode satellite (Kosugi et al. 2007) it is now possible to obtain spectropolarimetric 
data (full Stokes vector) with high spatial resolution (0.3") and low noise ($\sigma \approx 10^{-3}$ in units of the 
continuum intensity). Thanks to these new data, it is now also possible to investigate not only the module but the three components
of the magnetic field vector. This has led to a new controversy about the angular distribution of the magnetic field vector
in the quiet Sun. While some authors (Orozco Su\'arez et al. 2007a, 2007b; Lites et al. 2007, 2008) found that the magnetic field 
is mostly horizontal ($\gamma \approx 90\deg$; with $\gamma$ being the inclination of the magnetic field vector with respect to the observer's 
line-of-sight), others favor a quasi-isotropic distribution of magnetic fields (Mart{\'\i}nez Gonz\'alez et al. 2008; 
Asensio Ramos 2009; Stenflo 2010). With a few exceptions (Harvey et al. 2007, Lites et al. 2008 and Mart{\'\i}nez 
Gonz\'alez et al. 2008), all previous studies were carried out employing data recorded at disk center only. Therefore, 
to better constrain the angular distribution of the magnetic field vector in the internetwork, we considered spectropolarimetric
data recorded at different positions on the solar disk (Section ~\ref{section:observations}).\\

In addition, Asensio Ramos (2009), Stenflo (2010), and Borrero \& Kobel (2011; hereafter referred to as paper I) warned 
that the highly inclined magnetic fields obtained by some studies could be caused by the noise in the linear 
polarization profiles. This yields a distribution of $B_\perp$ (component of the magnetic field vector that is 
perpendicular to the observer's line-of-sight) with a peak at around 50-90 Gauss. To avoid this problem, these 
authors proposed to include only those profiles in the analysis that have a signal-to-noise ratio $> 4.5$ in the linear 
polarization (Stokes $Q$ and $U$). Although this selection criterion allows one to retrieve reliable distributions for 
the magnetic field vector, it has the disadvantage of excluding most of the Stokes profiles within the field-of-view from 
the analysis (see Borrero \& Kobel 2012; hereafter referred to as paper II; cf. Bellot Rubio \& Orozco Su\'arez 2012). 
In this paper we adopt an alternative approach based on inverting the histograms of the observed Stokes vector 
(Section~\ref{section:pdftheory}) over the entire field-of-view instead of inverting the Stokes vector at each pixel 
over the observed region. Under a number of simplifying assumptions, whose limitations are described in 
Section~\ref{section:limitations}, we were able to reach some important, albeit preliminary, conclusions about the
angular distribution of the magnetic field vector in the solar internetwork and its variation across the solar disk 
(Section~\ref{section:conclu}).\\

\section{Observations and datasets}
\label{section:observations}

The data employed in this work were recorded with the spectropolarimeter (SP; Ichimoto et al. 2008) attached
to the Solar Optical Telescope (SOT; Tsuneta et al. 2008, Suematsu et al. 2008, Shimuzu et al. 2008) onboard 
the Japanese spacecraft Hinode (Kosugi et al. 2007). The spectropolarimetric data comprise the full Stokes
vector $(I, Q, U, V)$ around the pair of magnetically sensitive spectral lines \ion{Fe}{I} 6301.5 {\AA} ($g_{\rm eff}=1.67$)
and \ion{Fe}{I} 6302.5 {\AA} ($g_{\rm eff}=2.5$). $g_{\rm eff}$ refers to the effective Land\'e factor calculated under
LS coupling. The spectral resolution of these observations is about 21.5 m{\AA} per pixel, with 112 pixels in the 
spectral direction. The spatial resolution of Hinode/SP observations is 0.32". For this paper we selected three 
maps at three different heliocentric positions. In all three maps the spectrograph's 
slit was kept at the same location on the solar surface for the whole duration of the scan. This means that, while the 
vertical direction ($Y$-axis or direction along the slit) contains information about different spatial structures on 
the solar surface, the horizontal direction ($X$-axis or direction perpendicular to the spectrograph's slit) samples the
 same position at different times. Each spectrum was recorded with a 9.6 seconds exposure, yielding a noise of about 
$\sigma = 7.5 \times 10^{-4}$ in units of the quiet-Sun continuum intensity. Each map records data for 
a period of time ($> 1$ hr) that includes several turnovers of the granulation, thus breaking down the temporal coherence
and providing spatial information (in a statistical sense) along the $X$-axis.\\

In paper I we have demonstrated that photon noise plays an important role in determining
the magnetic field vector from spectropolarimetric observations. To further decrease the level of noise in our
observations we averaged every seven slit positions (temporal average of about 67.1 seconds), which
yields a new noise level of about $\sigma = 3 \times 10^{-4}$ (in units of the quiet-Sun continuum intensity). 
However, averaging means that the original map is shortened by a factor of seven in the direction that is perpendicular 
to the slit ($X$-axis). This decreases the number of points available for statistics. Fortunately, Hinode/SP data have 
a sufficient number of pixels to ensure good statistics even after averaging (see Section \ref{section:conclu}). In 
the following we briefly describe each map individually.

\subsection{Map A}
\label{subsection:mapa}

This map was recorded on February 27, 2007 between 00:20 UT and 02:20 UT. It originally consists 
of 727 slits positions, of which 103 remain after temporal averaging. The center of slit was located at
approximately the following coordinates on the solar surface: $X = -31.7"$ and $Y = 7.7"$. This corresponds
to a heliocentric position of $\mu=\cos\Theta \approx 1$ ($\Theta$ is the heliocentric angle) and to a
latitude of $\Lambda \approx 0\deg$. The noise level is $\sigma = 2.8 \times 10^{-4}$. This map 
(original and temporally averaged) corresponds to Maps B and C in paper I, and it was also
employed (with and without temporal averaging) by Lites et al. (2008) and Orozco Su\'arez et al. (2007a).

\subsection{Map B}
\label{subsection:mapb}

This map was recorded on February 6, 2007 between 11:33 UT and 15:51 UT. It originally consists 
of 1545 slits positions, of which 222 remain after temporal averaging. The center of slit was located at
approximately the following coordinates on the solar surface: $X = 493.6"$ and $Y = 491.3"$. This corresponds
to a heliocentric position of $\mu=\cos\Theta \approx 0.7$ and to a latitude of $\Lambda \approx 30\deg$. The noise 
in this map is very similar to that in Map B: $\sigma = 3.1 \times 10^{-4}$.\\

\subsection{Map C}
\label{subsection:mapc}

This map was recorded on January 17, 2007 between 07:05 UT and 09:58 UT. It originally consists 
of 1048 slits positions, of which 149 remain after temporally averaging. The center of slit was located at
approximately the following coordinates on the solar surface: $X = -3.0"$ and $Y = 697.1"$. This corresponds
to a heliocentric position of $\mu=\cos\Theta \approx 0.7$ and to a latitude of $\Lambda \approx 40\deg$. Here, the 
noise level is slightly higher than in Map A: $\sigma = 3.2 \times 10^{-4}$. We note that some consecutive slit positions 
in this map show very high noise in the Stokes profiles. Although we could not relate this effect to the South Atlantic 
Anomaly (increased flux of cosmic rays at certain orbits of the satellite) we have removed these slit positions 
from our analysis, which reduced the effective number of slit positions to 120.\\

\begin{figure}
\begin{center}
\includegraphics[width=9cm]{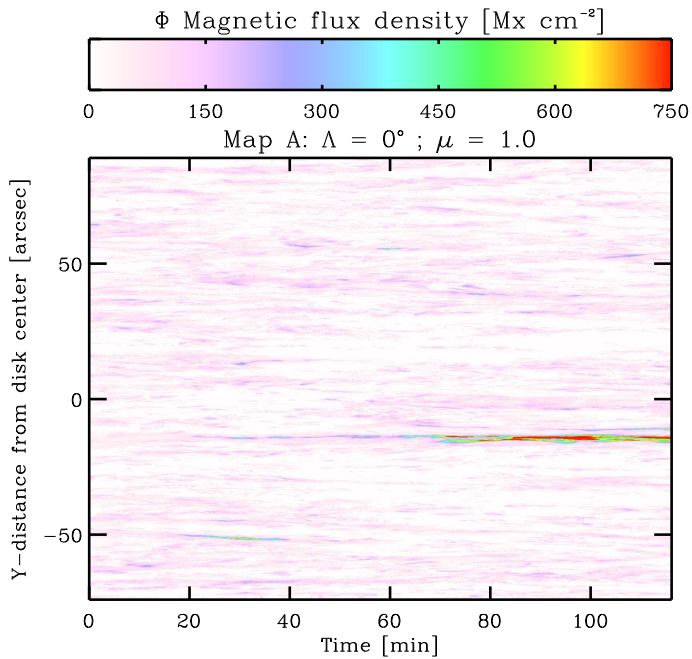}
\end{center}
\caption{Inferred magnetic flux density $\Phi=\alpha B$ from the inversion of Map A (Sect.~\ref{subsection:mapa}).
White areas correspond to regions where all three polarization profiles (Stokes $Q$, $U$, and $V$) are below
the $4.5\sigma$-level.}
\label{figure:invmapa}
\end{figure}

\begin{figure}
\begin{center}
\includegraphics[width=9cm]{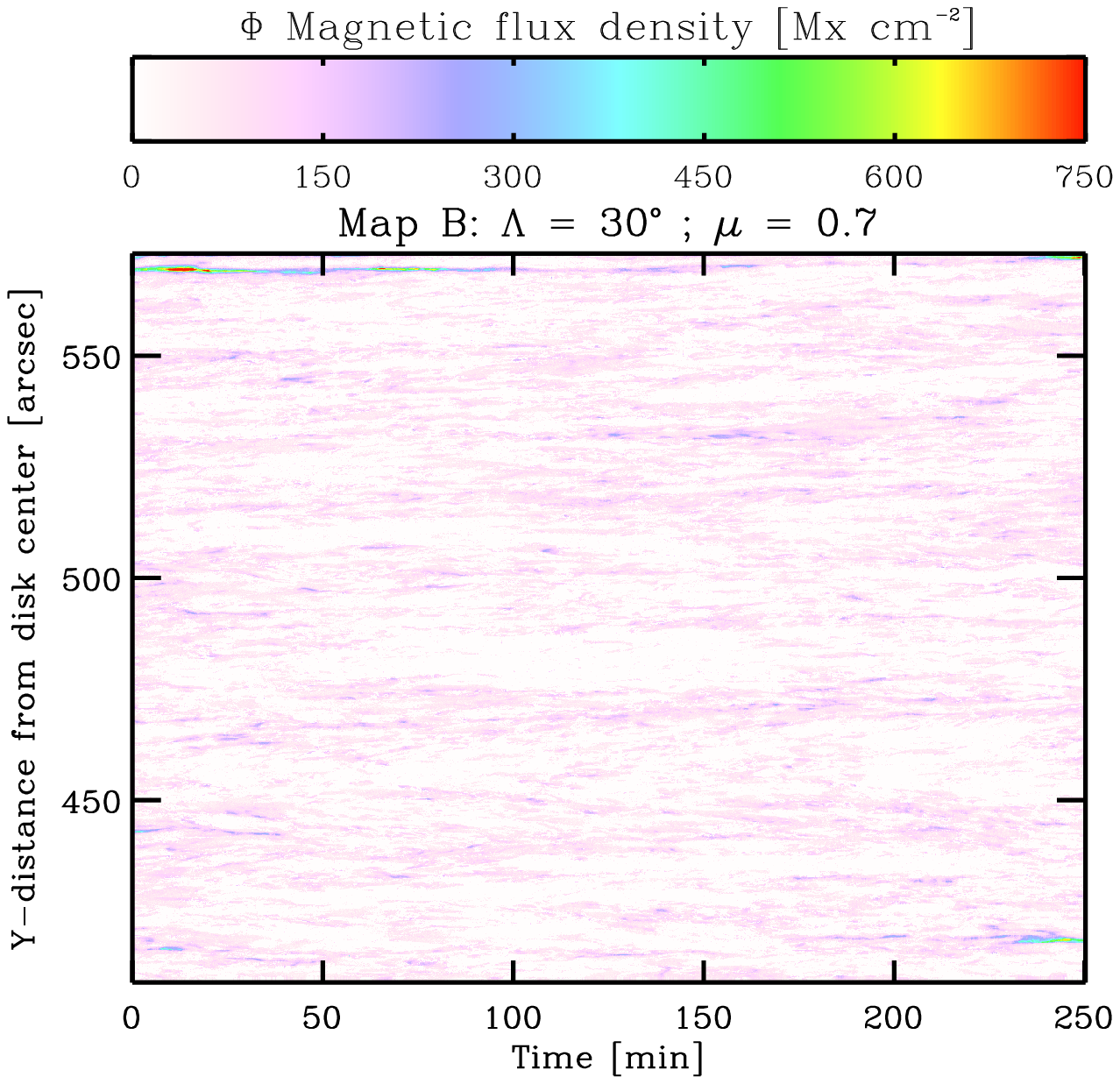}
\end{center}
\caption{Same as Figure~\ref{figure:invmapa} but for map B (Sect.~\ref{subsection:mapb}).}
\label{figure:invmapb}
\end{figure}

\begin{figure}
\begin{center}
\includegraphics[width=9cm]{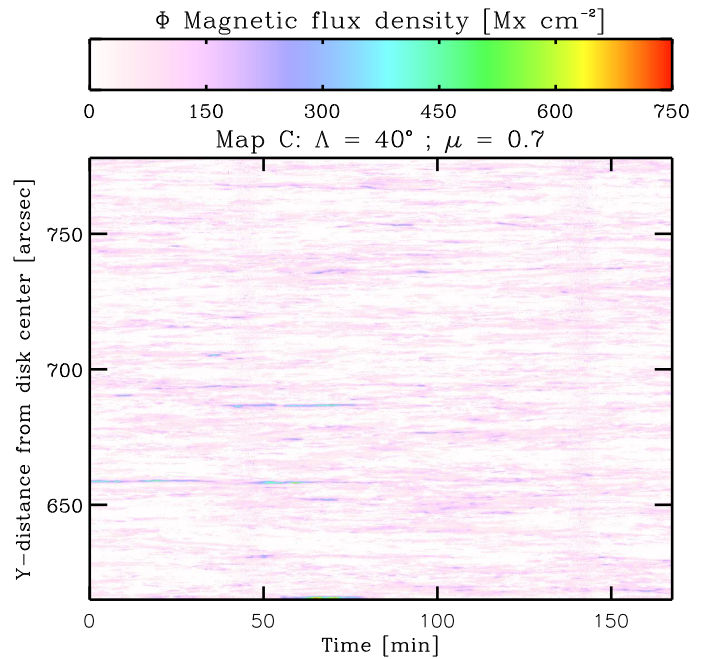}
\end{center}
\caption{Same as Figure~\ref{figure:invmapa} but for map C (Sect.~\ref{subsection:mapc}).}
\label{figure:invmapc}
\end{figure}

Figures \ref{figure:invmapa},~\ref{figure:invmapb}, and \ref{figure:invmapc} display the magnetic flux density $\Phi$
of maps A, B, and C as obtained through the inversion of the full Stokes vector with the VFISV 
(Very Fast Inversion of the Stokes Vector) inversion code (Borrero et al. 2010). For better visualization the maps 
in these figures are obtained from the inversion of the original data (i.e. not temporally averaged). This avoids 
pixelization in the $X$-axis of these plots. However, for the remainder of the paper, our discussions and 
figures are based only on the temporally averaged (67.1 seconds) data.\\

Although these previous figures only show the total magnetic flux density, it is worth mentioning that the VFISV code also retrieves the 
three components of magnetic field vector $\ve{B}$: $B$ is the module of $\ve{B}$, $\gamma$ is the inclination of $\ve{B}$ with respect
to the observer's line-of-sight, and $\phi$ is the azimuth of $\ve{B}$ in the plane that is perpendicular to the observer's line-of-sight.
In addition, VFISV retrieves the magnetic filling factor $\alpha$ as well as the line-of-sight component of the velocity vector 
$V_{\rm los}$ and a set of thermodynamic parameters $\ve{T}$. We note that the magnetic flux density is defined as $\Phi = \alpha B$.
For a more detailed overview on Milne-Eddington inversion codes, which include not only the magnetic field vector but also the thermodynamic 
and kinematic parameters relevant to the line formation, we refer the reader to del Toro Iniesta (2003), Borrero et al. (2010) and references 
therein.\\

\section{Stokes profiles at different positions on the solar disk.}
\label{section:clvobs}

\begin{figure*}
\begin{center}
\begin{tabular}{cc}
\includegraphics[width=9cm]{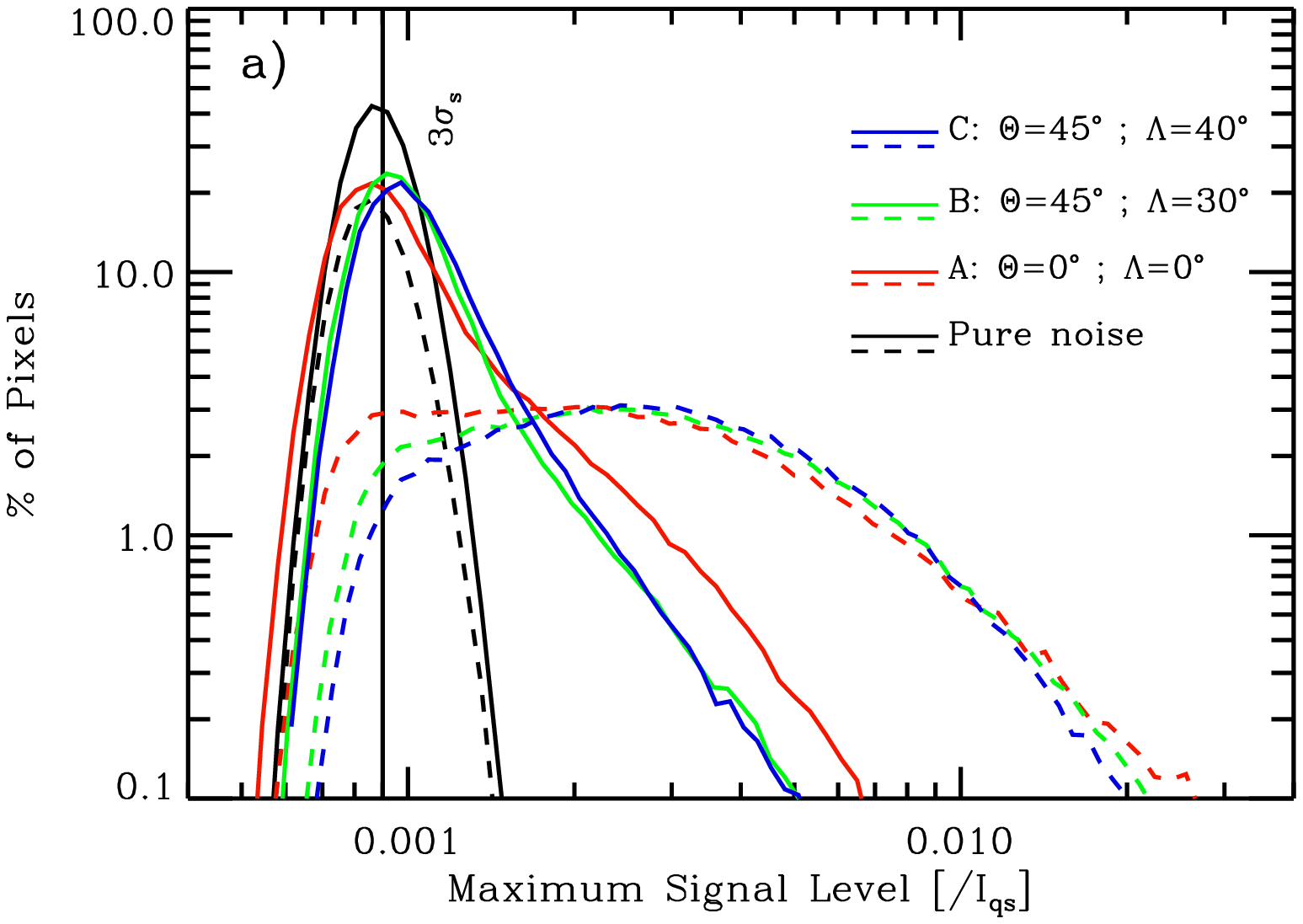} & 
\includegraphics[width=9cm]{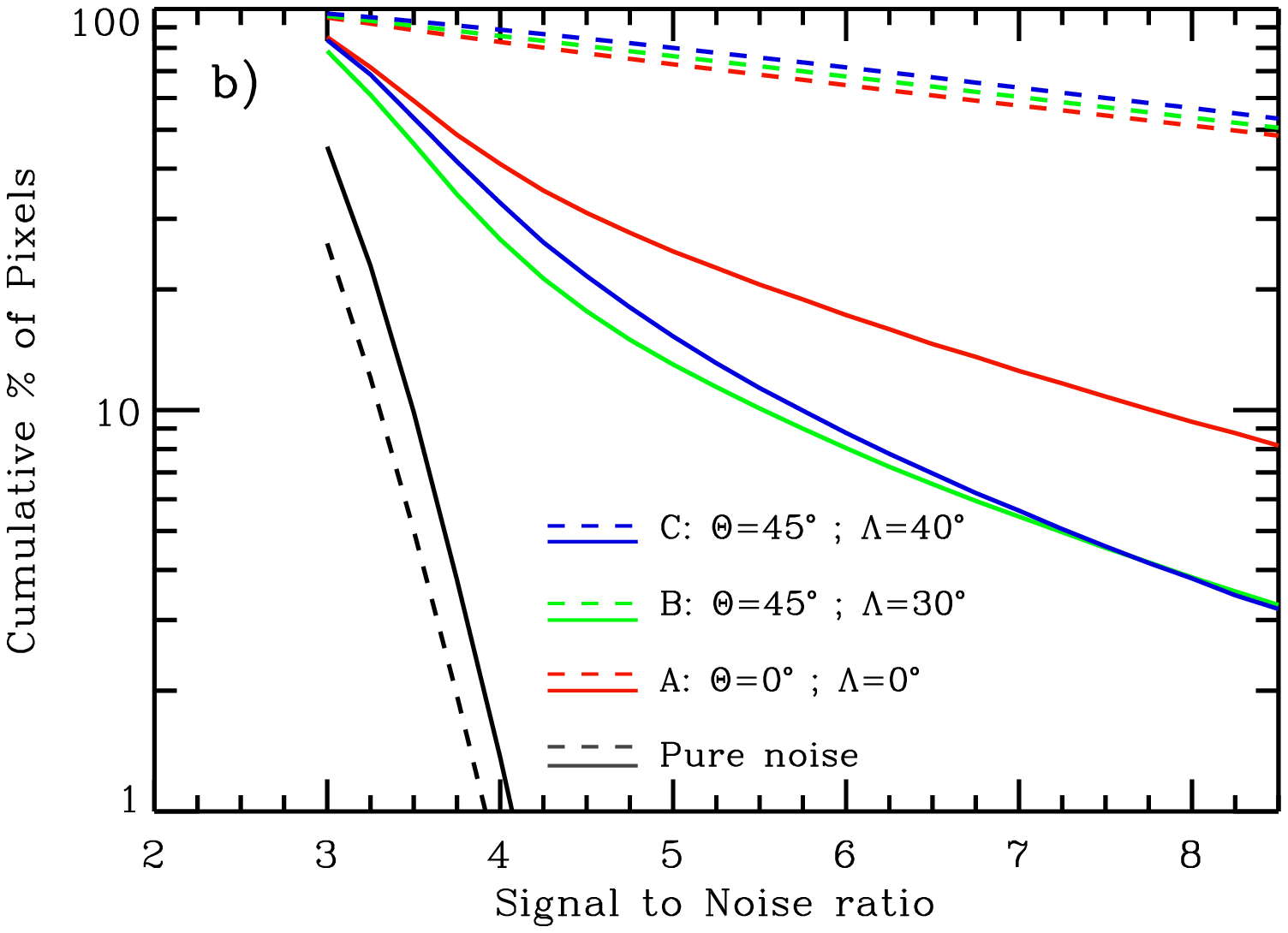} \\
\end{tabular}
\caption{{\it Left panel}: histogram of the number of pixels as a function of the maximum value in their polarization
signals  (normalized to the average continuum intensity). The vertical solid-black line represents the $3\sigma$-level.
{\it Right panel}: histogram of pixels with polarization signals above a certain signal-to-noise ratio ($S/R$). This 
last histogram is cumulative, while the first one is not. In both panels the color lines indicate the same:  red 
(map A at $\Lambda=0\deg$), green (map B at $\Lambda = 30\deg$), blue (map C at $\Lambda=40\deg$). The solid lines 
refer to the linear polarization profiles (Stokes $Q$ and $U$) while the dashed lines correspond to the circular polarization 
profiles (Stokes $V$). For comparison we also display, in solid and dashed-black curves, the expected behavior due to noise.}
\label{figure:stokhistogram}
\end{center}
\end{figure*}

The inversions carried out in the previous section could be employed to obtain histograms of the magnetic flux density $\Phi$,
module of the magnetic field vector $B$, and the inclination of the magnetic field vector with respect to the observer's 
line-of-sight ($\gamma$) at different positions on the solar disk. However, in this paper we aimed to infer properties about 
the distribution of the magnetic field vector by directly studying the histograms of the Stokes profiles. 
Figure~\ref{figure:stokhistogram}a presents distribution histograms of the maximum signals of the Stokes $V(\lambda)$ (dashed lines) 
and Stokes $Q(\lambda)$ and $U(\lambda)$ (solid lines) normalized to the average quiet-Sun intensity over the entire map: 
$I_{\rm qs}$. The colors indicate each of the different maps studied: red for map A (Sect.~\ref{subsection:mapa}), green for map B 
(Sect.~\ref{subsection:mapb}), and blue for map C (Sect.~\ref{subsection:mapc}). Figure~\ref{figure:stokhistogram}b displays 
the cumulative histogram of the pixels in each map that have a $S/R$ (signal-to-noise ratio) equal to or higher than a given value. 
The colors and the line-styles are as in Figure~\ref{figure:stokhistogram}a. For instance: 31.6 \% of the pixels in map A 
posses signals in $Q$ or $U$ (solid-red line) that are above 4.5 times the noise level. To limit our analysis to the internetwork 
regions, we excluded from these figures the pixels in maps A, B, and C with a magnetic flux density $\Phi > 500$ Mx cm$^{-2}$.\\

We now focus on some of the features of the histograms in Figure~\ref{figure:stokhistogram}. A very striking one is the peak
at around the $3\sigma$-value for the linear polarization (solid lines) in Fig.~\ref{figure:stokhistogram}a, and the fact
that the amount of pixels with maximum linear polarization signals below and above this $3\sigma$-level quickly drops. One 
might wonder how a peak can appear at around 3$\sigma$ if the probability that photon noise (taken as a normal random distribution) 
will produce such a high value is only about 0.27 \% ? The answer to this question is that photon noise has a probability of 0.27 \% 
to produce a signal at the 3$\sigma$-level at one particular wavelength, but Figure~\ref{figure:stokhistogram} shows the maximum 
of the signal over all wavelengths. Indeed, it is possible to employ the binomial distribution to find the lower bound of the 
probability $K(\delta,N)$ that after $N$ wavelengths, one of them will have a signal stronger than or equal to 
$\delta$-times the noise level $\sigma$:

\begin{equation}
K(\delta,N) \geq 1-[1-p(\delta)]^N \;,
\end{equation}

\noindent where $p(\delta)$ is the probability that at one single wavelength position, a normally
distributed random variable (with a standard deviation $\sigma$) will yield a signal $\delta$-times above the 
standard deviation is given by

\begin{equation}
p(\delta) = 1-\textrm{erf}(\delta/\sqrt{2}) \;,
\end{equation}

\noindent where $\textrm{erf}$ denotes the so-called error function. Since the spectral line is sampled in $N=112$ 
spectral positions for each Stokes parameter, the probability of finding a wavelength where the noise yields a signal
 at the 3$\sigma$ is much higher than the 0.27 \% mentioned above. In particular, there is a $K(N=112,\delta=3) \geq 26.12 \%$ 
probability that the noise in the circular polarization (Stokes $V$) will yield a signal at the $3\sigma$-level. Because 
the linear polarization consists of two Stokes parameters ($Q$ and $U$), this probability is even higher: $K(N=224,\delta=3) 
\geq 45.42 \%$. Given these high probabilities, it is not surprising that the histograms of the Stokes
profiles peak close to the $3\sigma$-level. This is certainly the case of the linear polarization (solid lines in 
Fig.~\ref{figure:stokhistogram}a). It is noteworthy that, at the $3\sigma$-location, Stokes $V$ (circular polarization) 
only presents a local maximum (dashed lines in  Figure~\ref{figure:stokhistogram}a) that actually disappears for maps B 
(dashed green) and C (dashed red). Indeed, the peak for the histogram of Stokes $V$ appears to be located in the range 
of $2-3\times 10^{-3}$ (in units of the quiet Sun continuum intensity), which is about 8 times above the noise level $\sigma$.\\

\begin{figure}
\begin{center}
\includegraphics[width=9cm]{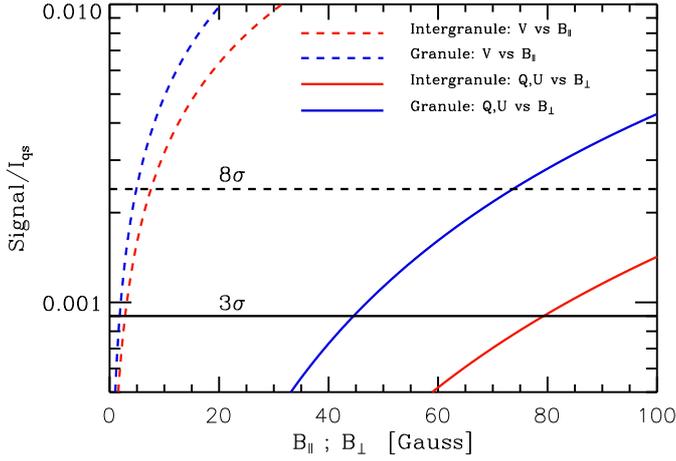}
\caption{Amount of circular polarization (normalized to the quiet-Sun continuum intensity) generated as a function of the component of 
the magnetic field vector that is parallel to the observer's line-of-sight: $B_\parallel$ (dashed lines). Amount of linear polarization 
(normalized to the quiet-Sun continuum intensity) signals generated as a function of the component of the magnetic field vector that 
is perpendicular to the observer's line-of-sight: $B_\perp$ (solid lines). These curves were obtained by performing a synthesis
of the Stokes profiles employing different values of $B_\parallel$ and $B_\perp$ and using two different atmospheric semi-empirical 
models: red for intergranules and blue for granules. The horizontal black lines indicate the $3\sigma$
and $8\sigma$ levels assuming a noise of $\sigma=3\times 10^{-4}$ as in our observed maps (Section~\ref{section:observations}).}
\label{figure:bfnoise}
\end{center}
\end{figure}

These differences between linear and circular polarization in Figure~\ref{figure:stokhistogram}a can be explained if we consider a 
distribution of $B_\parallel$ (component of the magnetic field vector that is parallel to the observer's line-of-sight) that 
features a peak at a value such that the corresponding Stokes $V$ profile would have a maximum at around $8\sigma$. At the same 
time, the distribution of $B_\perp$ (component of the magnetic field vector that is perpendicular to the observer's line-of-sight) 
should feature a low probability of finding values of $B_\perp$ that produce Stokes $Q$ and $U$ profiles above
$3\sigma$ such that there is a peak at this level. We additionally investigated this by synthesizing Stokes profiles with different 
values of $B_\parallel$ and $B_\perp$ and comparing them with the maximum signal of the resulting Stokes profiles. To obtain a 
good estimation we carried out this experiment with two different semi-empirical models that represent granules and intergranules 
(Borrero \& Bellot Rubio 2002). Results are presented in Figure~\ref{figure:bfnoise}. This figure shows that the distribution of 
$B_\parallel$ that is responsible for the Stokes $V$ signal in Figure~\ref{figure:stokhistogram}a (dashed-lines) must posses a 
peak at around $B_\parallel \approx 5-7$ G, which is the value needed to produce a majority of Stokes V signals at the 
$8\sigma$-level. In addition, Figure~\ref{figure:bfnoise} shows that the distribution of $B_\perp$ must have very low 
probabilities for $B_\perp \geq 40-70$ G, otherwise the peak in the histograms for $Q$ and $U$ in  Figure~\ref{figure:stokhistogram}a 
(solid lines) would be significantly shifted above the $3\sigma$-level.\\

An important feature of Figures~\ref{figure:stokhistogram}a and \ref{figure:stokhistogram}b is that the distribution
of the linear polarization signals (solid lines) is different at different positions on the solar surface. In addition, the distribution of 
the circular polarization signals (dashed lines) also changes slightly, but comparatively less than the linear polarization.\\

\section{Results from theoretical distributions}
\label{section:pdftheory}

In the previous section we have determined some general properties about the distribution of the magnetic field vector in the internetwork 
by considering some details from the observed histograms of the Stokes profiles. We now continue along these lines by investigating the 
sources of the differences in the distribution of the polarization signals at different positions on the solar disk. Following the 
notation introduced in paper II, we refer to $\ve{X}$ as the set of physical parameters that affect the formation of the Stokes profiles:\\

\begin{eqnarray}
\ve{X} = [\ve{T}, V_{\rm los}, \ve{B}, \alpha] \;,
\label{equation:x}
\end{eqnarray}

\noindent where $\ve{T}$ refers to the thermodynamic parameters, $V_{\rm los}$ to the line-of-sight-velocity, $\ve{B}$ to the 
magnetic field vector, and finally $\alpha$ refers to the magnetic filling factor (see also Section~\ref{section:observations}). 
Considering a Milne-Eddington atmosphere (see del Toro Iniesta 2003; Borrero et al. 2010) is equivalent to assuming that with 
the exception of source function, none of the thermodynamic, kinematic, and magnetic parameters change with optical depth $\tau_c$ 
in the photosphere. The source function, however, is considered to vary linearly with optical depth: $S(\tau_c) = S_0 +\tau_c S_1$, 
where $S_0$ corresponds to the source function at the observer's location, and $S_1$ is the derivative of the source function with 
optical depth. In addition to $S_0$ and $S_1$, the other thermodynamic parameters in $\ve{T}$ are the Doppler width of the spectral 
line $\Delta \lambda_{\rm D}$, the damping parameter $a$, and quotient of the  absorption coefficient in the continuum and in the 
line center $\eta_0$.\\

We now refer to $\mathcal{P}(\ve{X})\df \ve{X}$ as the probability of finding a pixel within the observed field-of-view where each 
of the physical parameters in $\ve{X}$ has values between $X_i$ and $X_i+\df X_i$. Furthermore, we assume that the magnetic parameters 
are statistically independent of the thermodynamic and kinematic parameters, thereby allowing us to write\\

\begin{eqnarray}
\mathcal{P}(\ve{X})\df \ve{X} = \mathcal{P}_1(\ve{B},\alpha) \mathcal{P}_2(\ve{T},V_{\rm los}) \df \ve{B} \df \ve{T} \df \alpha \df V_{\rm los} \;.
\label{equation:pdftot}
\end{eqnarray}

We now turn our attention to the probability distribution function of the magnetic parameters $\mathcal{P}_1$, which can be rewritten as\\

\begin{eqnarray}
\begin{split}
\mathcal{P}_1(\ve{B},\alpha)\df\ve{B} \df \alpha = & \mathcal{P}_1(B_i,B_j,B_k,\alpha) \df B_i \df B_j \df B_k \df \alpha \;,
\end{split}
\label{equation:pdfgeneric}
\end{eqnarray}

\noindent which indicates the probability of finding a pixel whose magnetic field vector has the coordinates between the 
following values: $B_i$ and $B_i + \df B_i$, $B_j$ and $B_j + \df B_j$, $B_k$ and $B_k + \df B_k$,
and finally where the filling factor of the magnetic field has a value between $\alpha$ and $\alpha + \df \alpha$. 
The reason for our choice of nomenclature in the three components of the magnetic field vector $B_i$, $B_j$ and $B_k$ is 
that the probability distribution function will generally not be expressed in spherical coordinates
in the observer's reference frame (the frame needed to solve the radiative transfer equation). In general,
we therefore must perform a transformation of variables that will express
Equation \ref{equation:pdfgeneric} into the observer's reference frame and into spherical coordinates:

\begin{eqnarray}
\begin{split}
\mathcal{P}_1(B_i,B_j,B_k,\alpha) \df B_i \df B_j \df B_k \df \alpha = &
|J| \mathcal{P}_1(B,\gamma,\phi,\alpha) \cdot \\ & \cdot \df B \df\gamma \df\phi \df \alpha \;,
\end{split}
\label{equation:coortransform}
\end{eqnarray}

\noindent where $|J|$ is the determinant of the Jacobian matrix for the transformation between the two
reference frames. This transformation might also introduce a dependence on the heliocentric angle
$\Theta$, thereby allowing us to evaluate the probability distribution function 
at different positions on the solar disk. The total probability must be equal to one:

\begin{equation}
\int_0^1 \df \alpha \int_0^\infty \df B \int_0^\pi \df\gamma \int_0^{2\pi} |J| \mathcal{P}_1(B,\gamma,\phi,\alpha)\df\phi = 1 \;.
\label{equation:normalization}
\end{equation}

We then use this distribution function to evaluate the percentage of pixels (from the total) that posses a given magnetic 
field vector. In our simulations we employed a total of $2\times 10^6$ pixels. To solve the radiative transfer equation, 
we need the probability distribution function of the thermodynamic and kinematic parameters $\mathcal{P}_2$ 
(see Eq.~\ref{equation:pdftot}), in addition to the probability distribution function of the magnetic parameters $\mathcal{P}_1$. 
Hereafter we take, as $\mathcal{P}_2(\ve{T},V_{\rm los}) {\textrm d}\ve{T} {\textrm d}V_{\rm los}$, the distribution
obtained from the results of map A (called map C in paper I). We assume that this distribution does not depend
with the position on the solar disk. Finally, in all our tests in this section we assume that the magnetic filling factor is unity: $\alpha=1$.
We take this approach not to add a new degree of freedom that will make our subsequent analysis more cumbersome.
Section~\ref{section:limitations} will address all the assumptions and simplifications of this section in more detail.\\

With this we now have all the necessary ingredients needed by the VFISV (Borrero et al. 2010) code to solve the radiative
transfer equation in order to obtain theoretical Stokes profiles $Q$, $U$, and $V$. To these profiles
we then add noise, assuming a normally distributed random variable (Leva 1992) with a standard deviation $\sigma = 3\times 10^{-4}$
(in units of the quiet-Sun continuum intensity). This noise level is similar to that of the observed maps in 
Section~\ref{section:observations}. Once noise has been added, we select the peak values of the Stokes profiles
and construct histograms like those derived from the observations in Figs.~\ref{figure:stokhistogram}a-\ref{figure:stokhistogram}b. 
The theoretical and observed histograms are then compared with different theoretical distributions of the magnetic field
in a attempt to match the observations.\\

The approach described in this section is indeed an inversion. However, it is not the same kind of inversion as those
in Section~\ref{section:observations} or in paper I. First of all, the observables here are the histograms of the peak 
values in Stokes $Q$, $U$, and $V$ (Figs.~\ref{figure:stokhistogram}a-\ref{figure:stokhistogram}b), whereas in paper I the observables were the
full Stokes vector ($I$, $Q$, $U$, and $V$) including their wavelength dependence (not only the peak values) at each individual pixel.
The model parameters also differ: while before the model parameters corresponded to the three components of the magnetic
field vector for each individual pixel, here the model parameters correspond to a parametrized theoretical distribution function
for the magnetic field vector that includes all pixels. As we show below, this parametrized theoretical distribution
function possesses a very limited number of free parameters (1-3), which can be tuned to simultaneously fit all pixels in the 
field-of-view and at different heliocentric angles. This is a great advantage compared to traditional inversion of Stokes profiles
(e.g. Borrero \& Kobel 2011; paper I), where there were about ten free parameters (those describing a Milne-Eddington atmosphere) for 
each pixel in the map. Another important difference is that, while regular inversion codes for the radiative transfer
equation (Ruiz Cobo \& del Toro Iniesta 1992; Borrero et al. 2010) are automatized, the procedure followed here is completely
manual (i.e: trial and error).\\

\subsection{Isotropic distribution functions}
\label{subsection:iso}

In this section we employ a theoretical distribution where the magnetic field vector is isotropic. We first
define as \emph{isotropic} a probability distribution function where the magnetic field vector has no preferred orientation.
In the local reference frame on the solar surface this can be expressed as having a probability of finding a magnetic field vector
that depends only on its module $B=\sqrt{B_{\rm x}^2+B_{\rm y}^2+B_{\rm z}^2}$:

\begin{equation}
\mathcal{P}_1(\ve{B})\df \ve{B} = A f(B)\df B_{\rm x} \df B_{\rm y} \df B_{\rm z} \;,
\label{equation:isocarte}
\end{equation}

\noindent where $A$ is just a normalization constant. In this reference frame, $\{\ex,\ey,\ez\}$, the $\ez$-axis is perpendicular
to the solar surface. To take into account that the observer's line-of-sight forms an angle $\Theta$ (heliocentric angle) with 
respect to the $\ez$-axis, we perform a variable change into a new coordinate system $\{\expr, \eypr, \ezpr \}$ that 
is rotated by an angle $\Theta$ around the $\ey$-axis. Indeed, because the distribution is isotropic, it does not matter around 
which axis we consider the rotation. With this transformation the new $\ezpr$-axis is aligned with the observer's 
line-of-sight. In this case, the relationship between the old coordinates of the magnetic field vector with the new ones is given by

\begin{eqnarray}
\begin{cases}
B_{\rm x} = B_{\rm z}^{'} \sin\Theta + B_{\rm x}^{'} \cos\Theta \;, \\
B_{\rm y} = B_{\rm y}^{'} \;, \\
B_{\rm z} = B_{\rm z}^{'} \cos\Theta - B_{\rm x}^{'} \sin\Theta \;, \\
|J| = 1 \;.
\end{cases}
\label{equation:rotation}
\end{eqnarray}

Since this is a simple rotation, the determinant of the Jacobian matrix for the transformation is unity. In addition, the module
of the magnetic field vector is the same in the old and new reference frames: $B=\sqrt{B_{\rm x}^2+B_{\rm y}^2+B_{\rm z}^2} =
\sqrt{B_{\rm x}^{'2}+B_{\rm y}^{'2}+B_{\rm z}^{'2}}$. We can therefore rewrite Equation~\ref{equation:isocarte} in the observer's reference 
frame as

\begin{equation}
 \mathcal{P}_1(\ve{B})\df \ve{B} = A f(B)\df B_{\rm x}^{'} \df B_{\rm y}^{'} \df B_{\rm z}^{'} \;,
\label{equation:isoobs}
\end{equation}

\noindent which is functionally identical to Eq.~\ref{equation:isocarte} since it was defined to be isotropic and therefore independent
of the viewing angle $\Theta$. As previously mentioned, to solve the radiative transfer equation we need to express Equation
~\ref{equation:isoobs} in spherical coordinates. We therefore perform now an additional variable change where

\begin{eqnarray}
\begin{cases}
B_{\rm x}^{'} = B\sin\gamma\cos\phi \;, \\
B_{\rm y}^{'} = B\sin\gamma\sin\phi \;, \\
B_{\rm z}^{'} = B\cos\gamma \;, \\
|J| = B^2\sin\gamma \;.
\end{cases}
\label{equation:spherical}
\end{eqnarray}

\noindent and thus,

\begin{equation}
\mathcal{P}_1(\ve{B})\df \ve{B} = A f(B) B^2 \sin\gamma \df B \df\gamma \df\phi \;.
\label{equation:iso}
\end{equation}

Since the distribution is isotropic, it must also be independent of $\Theta$ once it is expressed in the 
observer's reference frame both in Cartesian coordinates or spherical ones (Eqs.~\ref{equation:isoobs} and 
\ref{equation:iso}, respectively). In addition, isotropism manifests itself as a $\sin\gamma$-dependence, which 
comes from the determinant of the Jacobian matrix or, in order words, the volume-element expressed in spherical 
coordinates:  $\df\ve{B} = B^2 \sin\gamma \df B \df\gamma \df\phi$. For the distribution of the module of the magnetic 
field we consider an exponential function in the form $f(B) \approx \exp(-B)$. After normalization, the resulting expression for 
the theoretical distribution function is

\begin{equation}
\mathcal{P}_1(B,\gamma,\phi) \df B \df\gamma \df\phi = \frac{27}{8\pi}\frac{B^2}{B_0^3} \exp\left(\frac{-3B}{B_0}\right) \sin\gamma 
\df B \df\gamma \df\phi \;,
\label{equation:isotropic}
\end{equation}

\noindent where $B_0$ represents the mean value of the magnetic field vector module:

\begin{equation}
B_0 = <B> = \int_0^\infty \int_0^\pi \int_0^{2\pi}  B \cdot \mathcal{P}_1(B,\gamma,\phi) \df B \df\gamma \df\phi \;.
\end{equation}

\begin{figure*}
\begin{center}
\begin{tabular}{cc}
\includegraphics[width=9cm]{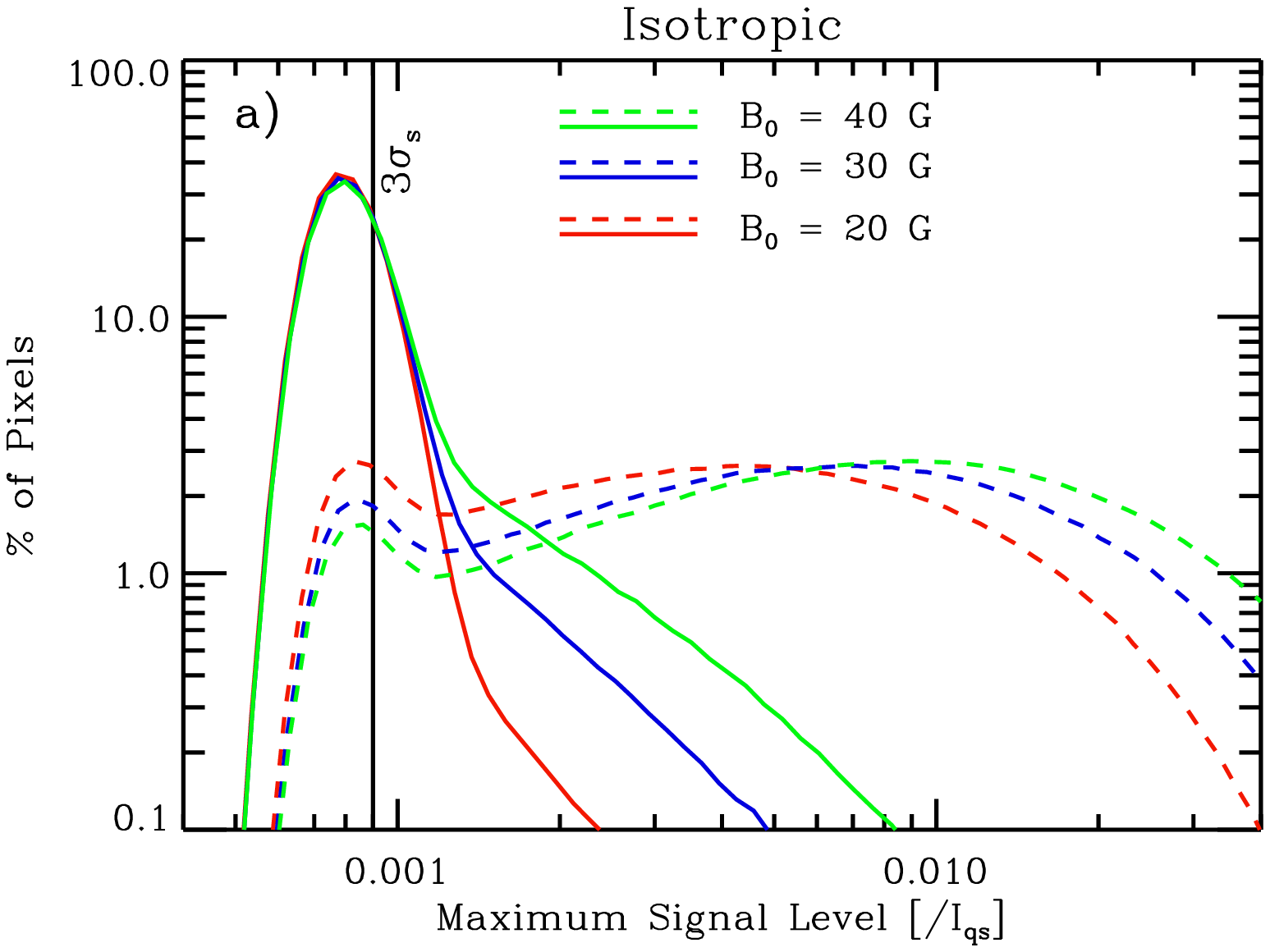} &
\includegraphics[width=9cm]{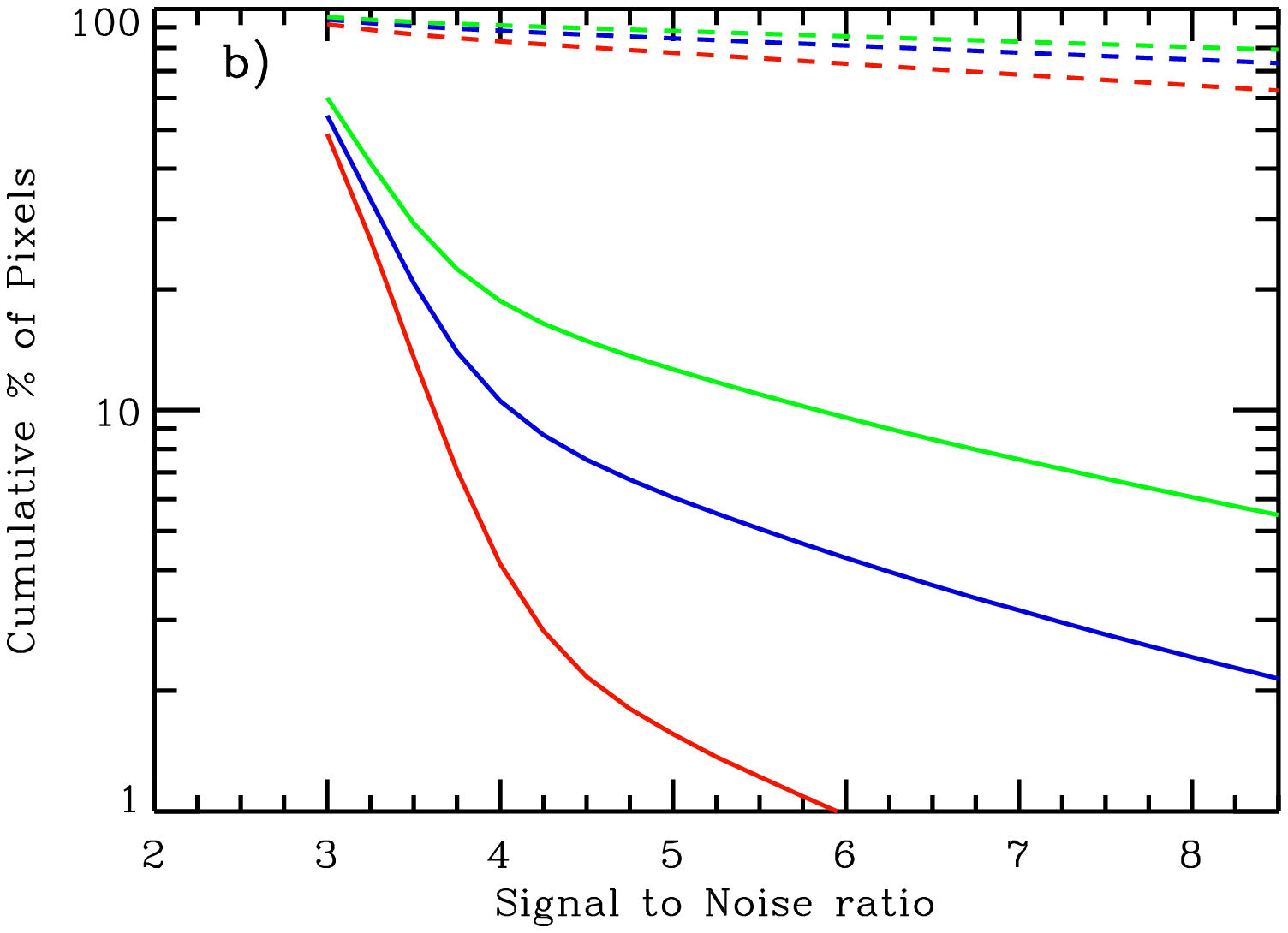} \\
\includegraphics[width=9cm]{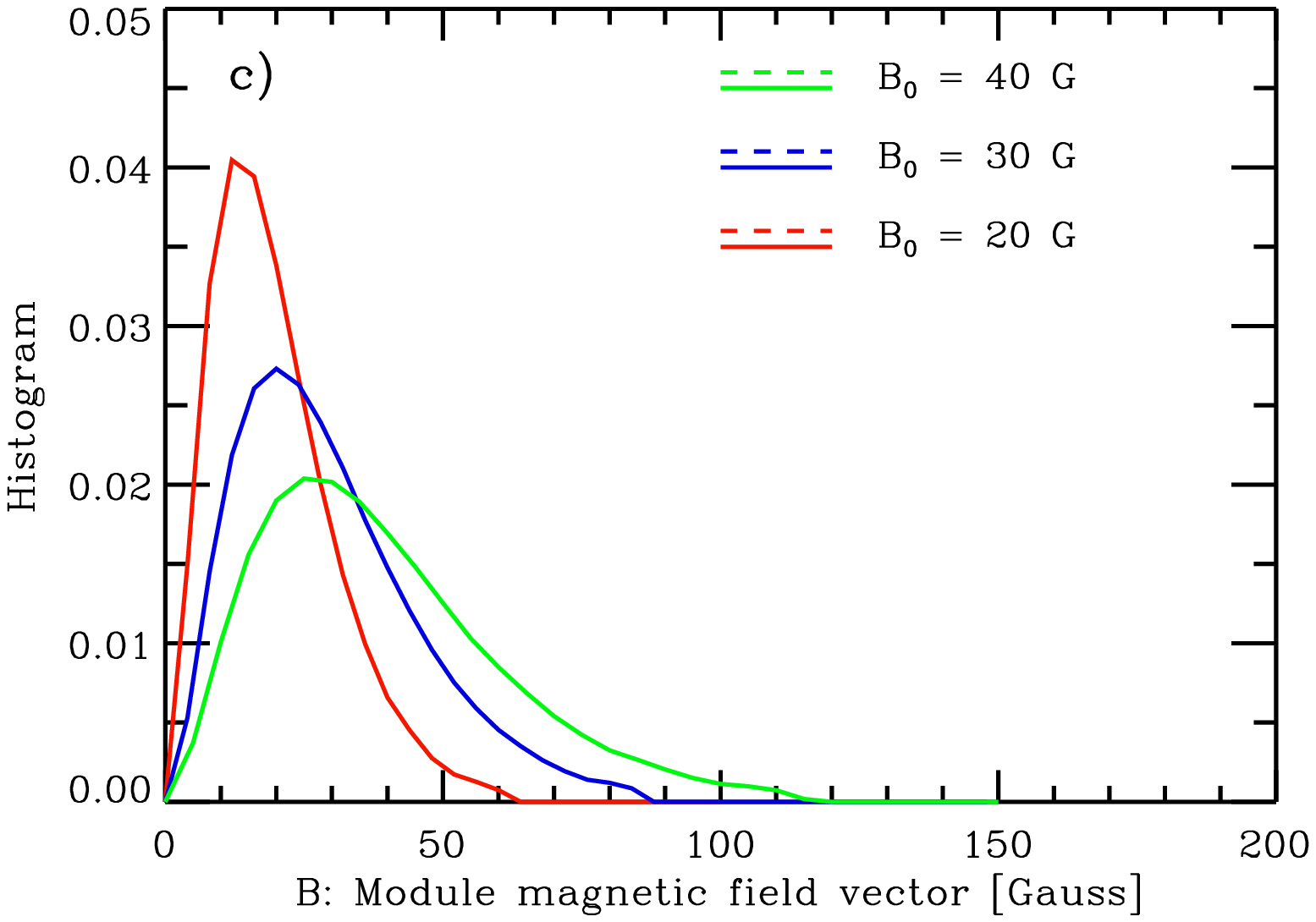} &
\includegraphics[width=9cm]{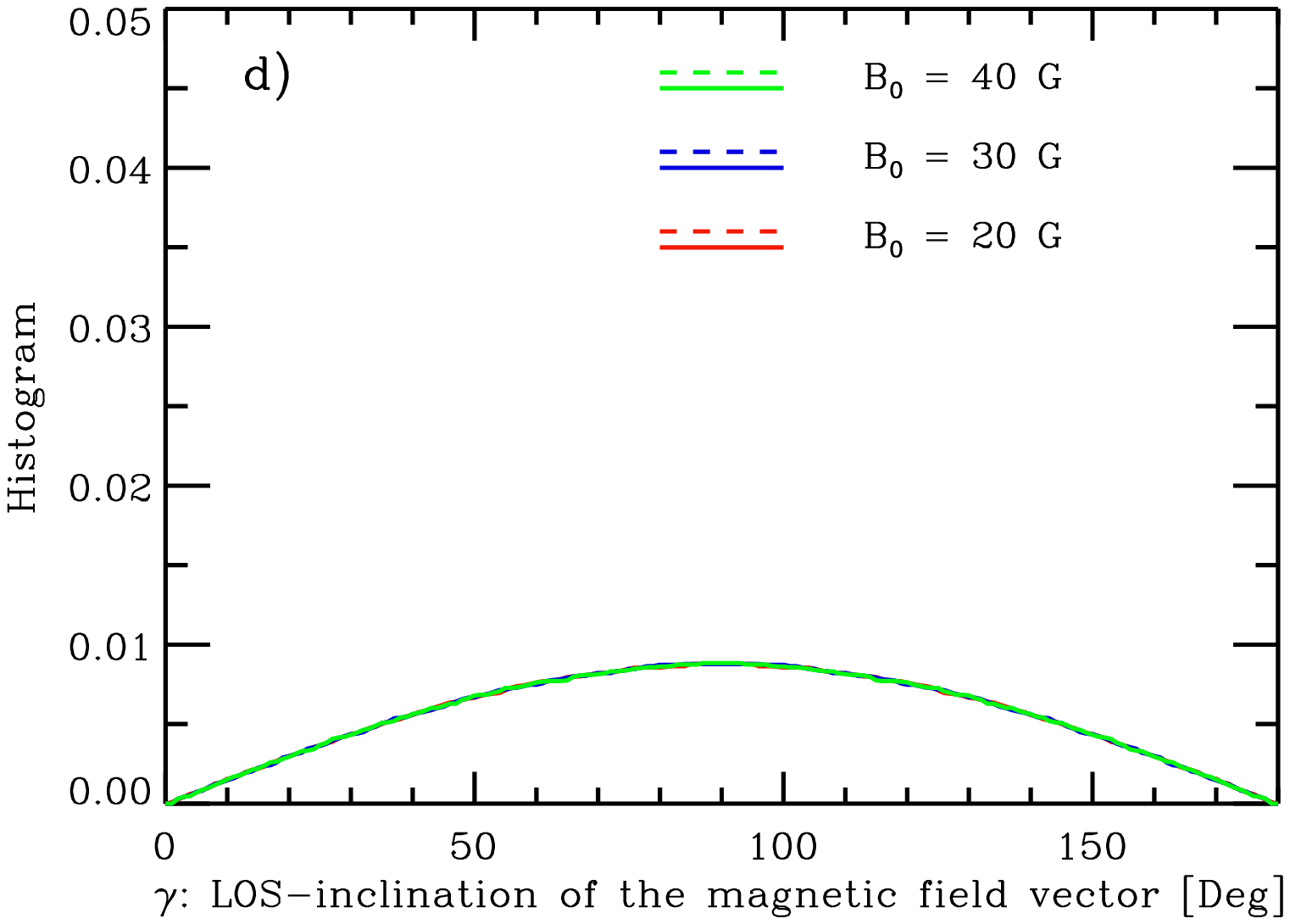}
\end{tabular}
\caption{{\it Panel a}: histogram of the number of pixels with a given maximum value in their polarization
signals (normalized to the average continuum intensity in the quiet Sun). The vertical solid-black line
corresponds to the $3\sigma$-level. {\it Panel b}: histogram of pixels with polarization signals above a certain signal-to-noise 
ratio ($S/R$). This last histogram is cumulative, but the first one is not. The solid and dashed lines refer to the 
linear and circular polarization profiles, respectively. These two panels can be readily compared to the observed histograms 
in Figure~\ref{figure:stokhistogram}. They were obtained employing an isotropic distribution function for the magnetic field 
vector (Eq.~\ref{equation:isotropic}) with three different mean values for the magnetic field strength $B_0$: 20 (red), 30 (blue),
 and 40 (green) Gauss. The module of the magnetic field vector $B$ and the inclination of this vector with respect to the observer's line-of-sight 
$\gamma$ arising from these isotropic distributions are presented in panels {\it c} and {\it d}.}
\label{figure:isotropic}
\end{center}
\end{figure*}

Figures~\ref{figure:isotropic}c-\ref{figure:isotropic}d display the probability distribution functions for module $B$ and inclination 
of the magnetic field vector with respect to the observer's line-of-sight $\gamma$, corresponding  to Eq.~\ref{equation:isotropic}, and 
employing three different values of $B_0 = 20, 30,$ and $40$ G. The histograms for the resulting Stokes profiles after solving the radiative transfer
equation with these distributions are given in Figures~\ref{figure:isotropic}a-\ref{figure:isotropic}b, which can be readily compared
 to Figs.~\ref{figure:stokhistogram}a-\ref{figure:stokhistogram}b. This comparison shows that for $B_0=20$ G, the linear polarization 
produced by an isotropic distribution (solid lines) is too low compared to the observed one. For $B_0=40$ G, the amount of linear polarization 
is comparable to the observed one, but, in this case the isotropic distribution produces too much circular polarization (dashed lines). 
In addition, an isotropic distribution produces far too few $Q$ and $U$ profiles, about 50 \% of the total, with $S/R > 3$ (solid lines in 
Fig.~\ref{figure:isotropic}b), while the observed distribution shows that about 70-80 \% of the profiles are above this level (solid lines in 
Fig.~\ref{figure:stokhistogram}b). The misfit between the theoretical histograms and the observed ones becomes even clearer when we consider 
that all curves in Fig.~\ref{figure:isotropic} are independent of the position on the solar disk (as they should, because they correspond to 
an isotropic distribution; Eq.~\ref{equation:isotropic}), while the observed ones (Fig.~\ref{figure:stokhistogram}) do change. Choosing a 
different $f(B)$ function in Eq.~\ref{equation:iso} does not alter this because the histograms of the Stokes profiles would still be 
independent of the position on the solar disk. However, it is possible to obtain histograms that vary with the position on the solar disk 
by postulating that the mean magnetic field depends upon $B_0(\Theta)$ or $B_0(\Lambda)$. In this case, the different curves in 
Fig.~\ref{figure:isotropic} will certainly change with the position on the solar disk, yet, in doing so we would implicitly introduce a 
dependence with $\Theta$ and/or $\Lambda$ in the original distribution, given by Eq.~\ref{equation:isotropic}, and thus it would no longer 
be isotropic.\\

\subsection{Triple-Gaussian distribution functions}
\label{subsection:triple}

We now consider a distribution function in the local reference frame on the solar surface, $\{\ex, \ey, \ez\}$, in 
which each component  of the magnetic field is statistically independent of the rest and shows a Gaussian-like 
dependence:

\begin{eqnarray}
\begin{split}
\mathcal{P}_1 (\ve{B})\df \ve{B} & = \frac{\df B_{\rm x} \df B_{\rm y} \df B_{\rm z}}{\pi^3 B_{\rm x0}B_{\rm y0} B_{\rm z0}}
\exp{\left\{-\frac{B_{\rm x}^2}{\pi B_{\rm x0}^2}\right\}} \cdot \\ & \cdot \exp{\left\{-\frac{B_{\rm y}^2}{\pi B_{\rm y0}^2}\right\}} \cdot
\exp{\left\{-\frac{B_{\rm z}^2}{\pi B_{\rm z0}^2}\right\}} \;.
\end{split}
\label{equation:triplelocal}
\end{eqnarray}

Again, with this definition, the $\ez$-axis is perpendicular to the solar surface. The distribution is of course normalized to one:

\begin{equation}
\iiint_{-\infty}^{\infty} \mathcal{P}_1(B_{\rm x},B_{\rm y},B_{\rm z}) \df B_{\rm x} \df B_{\rm y} \df B_{\rm z} = 1 \;.
\label{equation:norma}
\end{equation}

\begin{figure*}
\begin{center}
\begin{tabular}{cc}
\includegraphics[width=9cm]{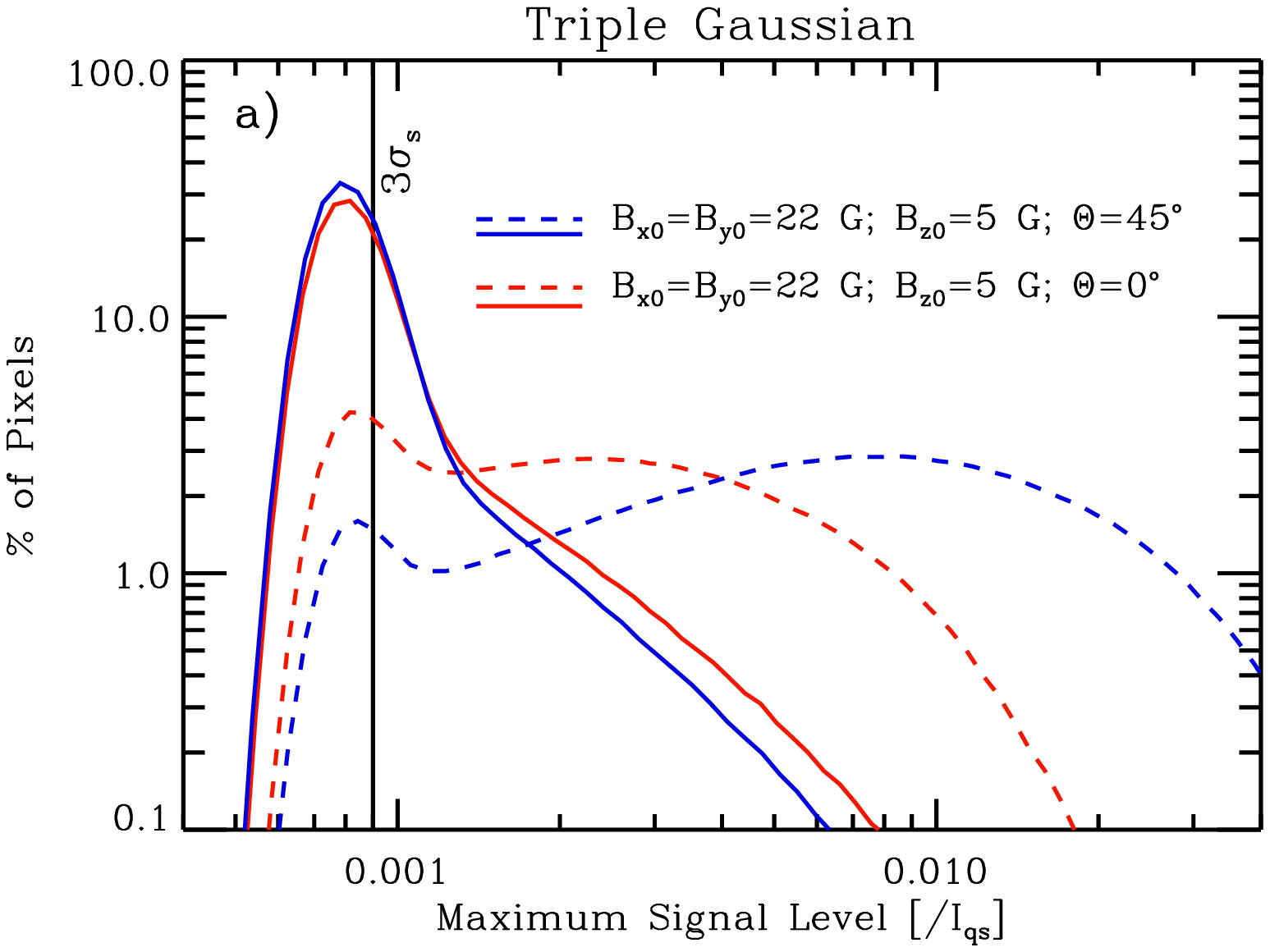} &
\includegraphics[width=9cm]{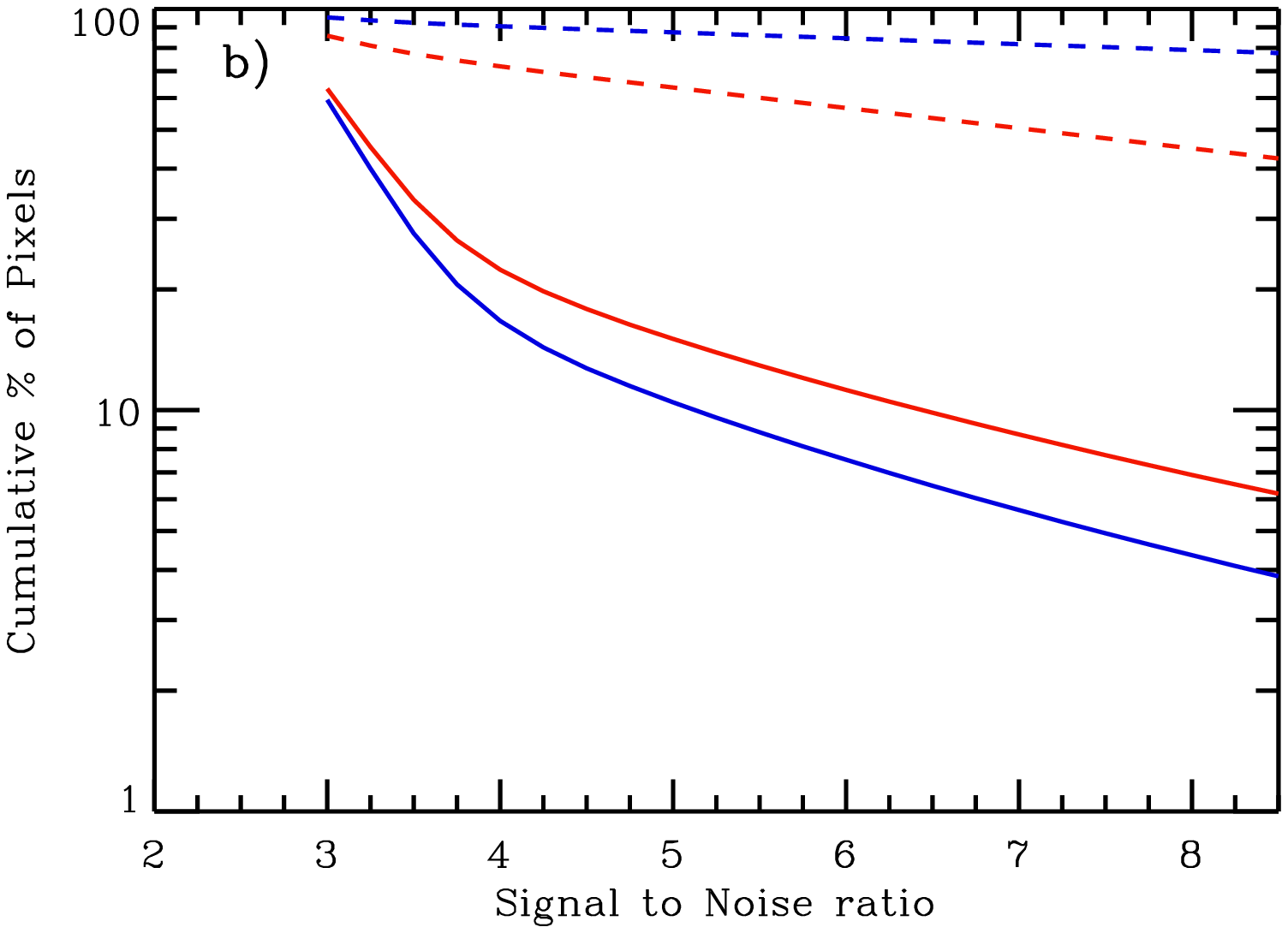} \\
\includegraphics[width=9cm]{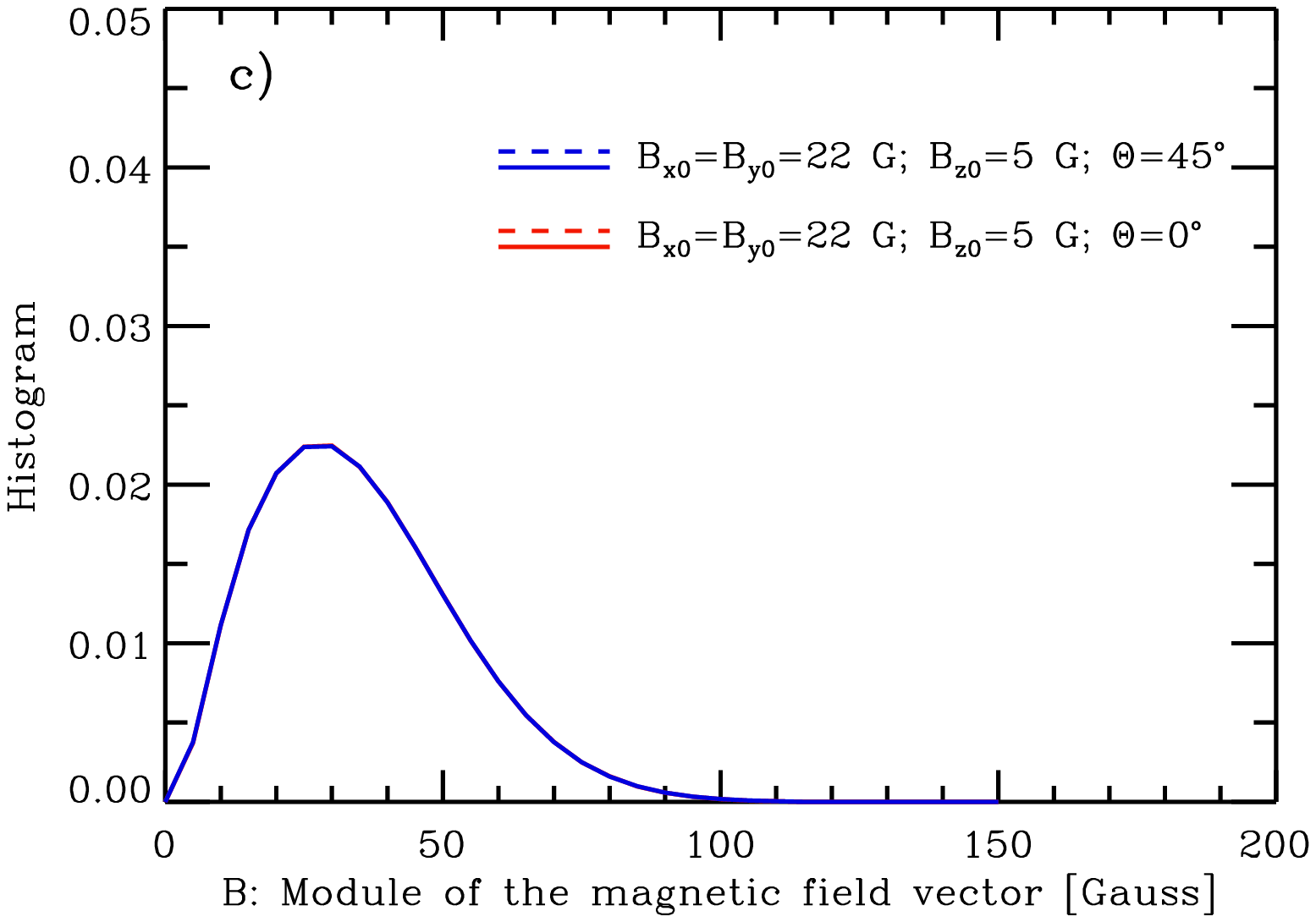} &
\includegraphics[width=9cm]{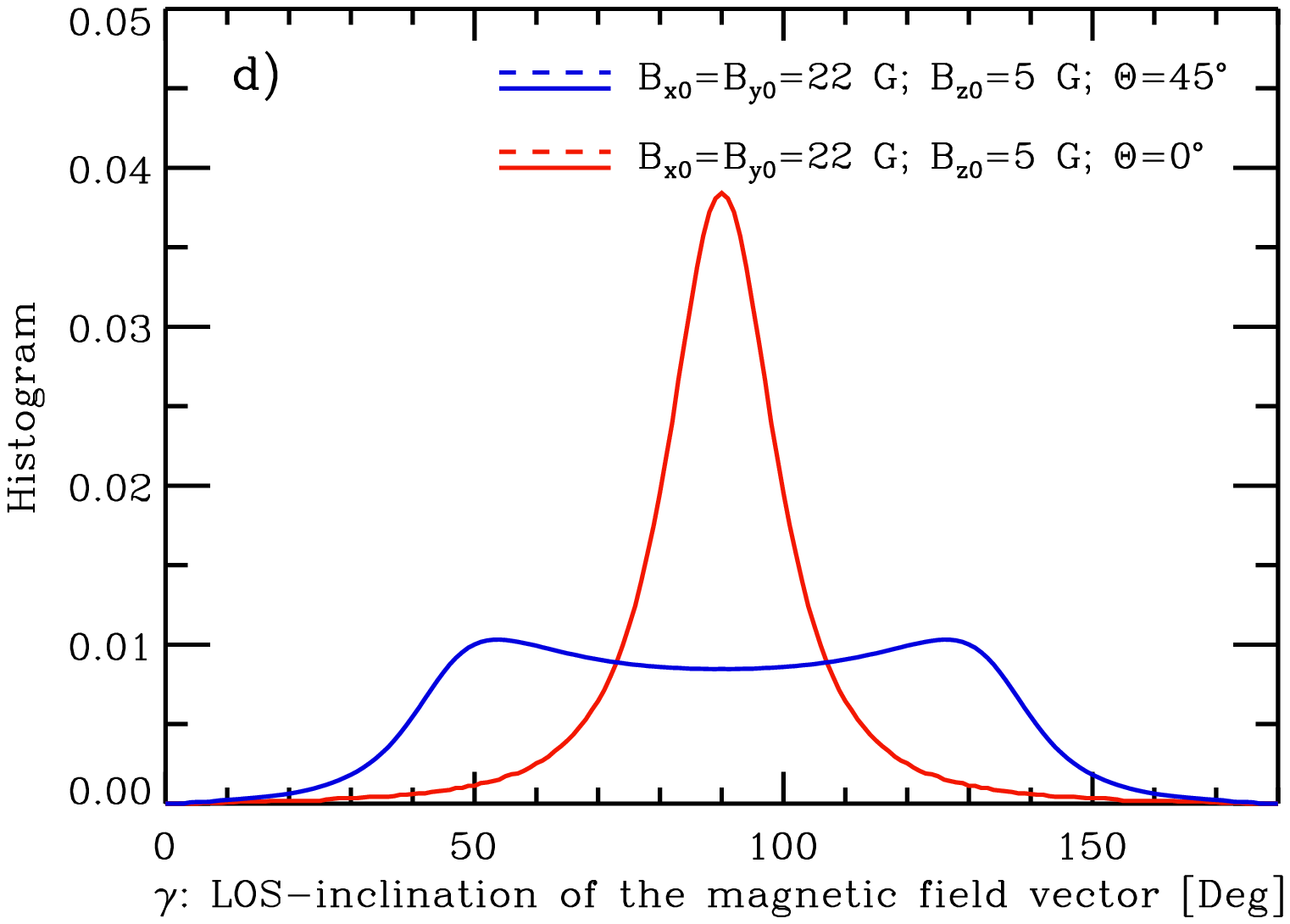}
\end{tabular}
\caption{Same as Figure~\ref{figure:isotropic} but for the \emph{triple-Gaussian} theoretical distribution given by Eq.~\ref{equation:triplesphe}. In
this case the mean value of the three components of the magnetic field vector are different: $B_{\rm x0}=B_{\rm y0}=22$ G, and $B_{\rm z0}=5$ G. Red lines 
correspond to $\Theta=0\deg$ (disk center) while green and blue lines correspond to $\Theta=45\deg$. Solid and dashed lines indicate 
the circular (Stokes $V$) and linear (Stokes $Q$ and $U$) polarization.}
\label{figure:triplegauss}
\end{center}
\end{figure*}

In the previous equation, $B_{\rm x0}$,  $B_{\rm y0}$, and  $B_{\rm z0}$ refer to the mean of the absolute value
for each component of the magnetic field:

\begin{equation}
B_{\rm x0} = \iiint_{-\infty}^{+\infty}
| B_{\rm x} | \mathcal{P}_1 (B_{\rm x},B_{\rm y},B_{\rm z})\df B_{\rm x} \df B_{\rm y} \df B_{\rm z} \;.
\end{equation}

Similar equations can be written for $B_{\rm y0}$ and $B_{\rm z0}$. As in the previous 
section, we perform a rotation of angle $\Theta$ around the $\ey$-axis to transform Eq.~\ref{equation:triplelocal} into the observer's 
reference frame: $\{\expr, \eypr, \ezpr \}$. To guarantee that the result of this transformation does not depend on the
choice of rotation axis in the $XY$-plane we must ensure that $B_{\rm x0}=B_{\rm y0}$ such that Eq.~\ref{equation:triplelocal} can be
written in terms of $B_{\rm x}^2+B_{\rm y}^2$. With this, we can now rewrite the theoretical distribution function as

\begin{eqnarray}
\begin{split}
\mathcal{P}_1 & (\ve{B})\df \ve{B} = 
\frac{\df B_{\rm x}^{'} \df B_{\rm y}^{'} \df B_{\rm z}^{'}}{\pi^3 B_{\rm x0}B_{\rm y0} B_{\rm z0}} 
\exp{\left\{-\frac{B_{\rm y}^{'2}}{\pi B_{\rm y0}^2}\right\}}\cdot \\
& \cdot \exp{\left\{-\frac{(B_{\rm z}^{'} \sin\Theta + B_{\rm x}^{'} \cos\Theta)^2}{\pi B_{\rm x0}^2}\right\}} \cdot \\
& \cdot \exp{\left\{-\frac{(B_{\rm z}^{'} \cos\Theta - B_{\rm x}^{'} \sin\Theta)^2}{\pi B_{\rm z0}^2}\right\}} \;.
\end{split}
\label{equation:tripleobs}
\end{eqnarray}

And finally, to solve the radiative transfer equation we need to express 
Equation~\ref{equation:tripleobs} in spherical coordinates (Eq.~\ref{equation:spherical}).
This transformation yields

\begin{eqnarray}
\begin{split}
\mathcal{P}_1 & (B,\gamma,\phi)\df B \df\gamma\df\phi =\frac{B^2 \sin\gamma\df B\df\gamma\df\phi}{\pi^3 B_{\rm x0}B_{\rm y0} B_{\rm z0}} \cdot \\
& \cdot \exp{\left\{-\frac{(B\sin\gamma\sin\phi)^2}{\pi B_{\rm y0}^2}\right\}} \cdot\\
& \cdot \exp{\left\{-\frac{(B\cos\gamma\sin\Theta+B\sin\gamma\cos\phi\cos\Theta)^2}{\pi B_{\rm x0}^2}\right\}} \cdot \\
& \cdot \exp{\left\{-\frac{(B\cos\gamma\cos\Theta-B\sin\gamma\cos\phi\sin\Theta)^2}{\pi B_{\rm z0}^2}\right\}} \;.
\end{split}
\label{equation:triplesphe}
\end{eqnarray}

With this theoretical distribution function of the magnetic field vector we can again (see Section~\ref{subsection:iso}) 
synthesize the Stokes profiles, add noise, and finally construct the theoretical histograms of the resulting Stokes profiles. 
Equations \ref{equation:tripleobs} and \ref{equation:triplesphe} now entail an explicit dependence on $\Theta$, unlike
the case of an isotropic distribution (Eq.~\ref{equation:isotropic}). Interestingly, if $B_{\rm x0}=B_{\rm y0}=B_{\rm z0}$, 
the distribution becomes isotropic, because Eq.~\ref{equation:triplelocal} can be re-written in terms of the module of the magnetic field 
vector $B$ (see Eq.~\ref{equation:isocarte}). In the following we employ $B_{\rm x0}=B_{\rm y0}=22 G; B_{\rm z0}=5$ G, and therefore 
the resulting distribution is non-isotropic. However, as indicated above, we impose $B_{\rm x0}=B_{\rm y0}$ so that the distribution is
symmetric in the $XY$-plane. The distributions for $B$ and $\gamma$ corresponding to the aforementioned values
are presented in Figures~\ref{figure:triplegauss}c and \ref{figure:triplegauss}d for two different heliocentric angles: $\Theta=0\deg$ 
(red curves), $\Theta=45\deg$ (blue). The reason for employing $B_{\rm z0} < B_{\rm x0},B_{\rm y0}$ is because in
the isotropic case (see Fig.~\ref{figure:isotropic}a) the value of $B_0$ that produced a reasonable fit to the linear 
polarization at disk center yielded too much circular polarization and thus, to simultaneously fit both, we need to decrease 
the vertical component of the magnetic field vector while keeping the horizontal component at the appropriate level.\\

The first feature to notice is that the distribution of $B$ (Fig.~\ref{figure:triplegauss}c) is independent of the heliocentric 
angle $\Theta$. This was to be expected because the module of the magnetic field vector $\ve{B}$ is invariant with respect to 
rotations. The second feature to notice is that the distribution of $\gamma$ shows a peak at $\gamma=90\deg$ at disk center 
(red curve). This is a consequence of having imposed $B_{\rm z0} < B_{\rm x0},B_{\rm y0}$. This peak smoothes out as we move 
toward the poles (blue curves).\\

The theoretical histograms for the Stokes profiles resulting from the previous distribution (Eq.~\ref{equation:triplesphe}) are 
displayed in Figures~\ref{figure:triplegauss}a and \ref{figure:triplegauss}b. As mentioned above, the values of $B_{\rm x0}$, $B_{\rm y0}$, 
and $B_{\rm z0}$ were selected such that, for signals above $2\times 10^{-3}$, the theoretical histograms at disk center 
($\Theta=0\deg$; red curves) are comparable to the observed ones (Fig.~\ref{figure:stokhistogram}). However, at larger heliocentric 
angles ($\Theta=45\deg$; blue curves), the mismatch between theoretical and observed histograms is very clear, in particular for the
 circular polarization (dashed lines). Our theoretical distribution function (Eq.~\ref{equation:triplesphe}) clearly predicts that 
the circular polarization should increase as $\Theta$ increases (blue dashed lines in Fig.~\ref{figure:triplegauss}). This happens 
as a consequence of having imposed $B_{\rm x0} > B_{\rm z0}$, which means that away from disk center, the component of the magnetic 
field vector that is aligned with the observer's line-of-sight, $B_{\rm z}^{'}=B_\parallel$, increases
due to the contribution from $B_{\rm x}$ (see Eq.~\ref{equation:rotation}). Because the slope of the $V-B_\parallel$ curve is so steep
(see Fig.~\ref{figure:bfnoise}), a small increase in $B_{\rm z}^{'}=B_\parallel$ translates into a large increase in the Stokes $V$ signal. 
Therefore, the generated amount of circular polarization is much larger at larger heliocentric angles (dashed-blue curves; $\Theta=45\deg$) 
than at disk center (dashed-red curves; $\Theta=0\deg$). Likewise, since $B_{\rm z0} < B_{\rm x0}$, the component of the magnetic field 
that is perpendicular to the observer's line-of-sight $B_\perp$ decreases as $\Theta$ increases. In this case, however, the linear 
polarization does not decrease as much as the circular polarization increases. This is due to the gentler slope of the $Q-B_\perp$ 
curve (Fig.~\ref{figure:bfnoise}).\\

Interestingly, Figs.\ref{figure:stokhistogram}a-\ref{figure:stokhistogram}b show that the observed amount of circular polarization 
does not increase, as required by the change in the viewing angle, for larger heliocentric angles (i.e. as $\Theta$ increases) in 
accordance to the observed drop in linear polarization. We conclude therefore that the differences in the histograms of the observed 
Stokes profiles at different positions on the solar disk cannot be produced by the change in the viewing angle $\Theta$. This problem 
is not unique to the theoretical distribution given by Eq.~\ref{equation:triplesphe}, but it will indeed affect any theoretical 
distribution that is prescribed in the local reference frame and features $B_{\rm z0} < B_{\rm x0},B_{\rm y0}$. Prescribing a 
theoretical distribution in the local reference frame means that the underlying distribution is always the same regardless of the 
position on the solar disk. The only reason that this distribution changes is because the angle $\Theta$, between the observer's 
line of sight and the vector perpendicular to the solar surface, varies with the position on the solar disk.\\

\subsection{Other distribution functions}
\label{subsection:other}

To avoid the problem described in Sect.~\ref{subsection:triple} we now prescribe a theoretical distribution of the magnetic 
field vector that, already in the local reference frame, depends on the latitude $\Lambda$. This will imply, unlike the 
previous distributions in Sections~\ref{subsection:iso} and \ref{subsection:triple}, that the underlying distribution 
(i.e. in the local reference frame) is different at different positions on the solar disk. In particular, the $\Lambda$ 
dependence means that the distribution function varies toward the poles, but not toward the limbs. The expression chosen 
in this section is\\

\begin{eqnarray}
\begin{split}
& \mathcal{P}_1(\ve{B}; \Lambda)\df \ve{B} = \frac{\df B_x \df B_y \df B_z}{192\pi\beta_\perp^5\beta_\parallel^3} (1+\zeta\sin\Lambda)^5 \cdot \\
& \cdot [(B_x\cos\Lambda+B_z\sin\Lambda)^2+B_y^2]^{\frac{3}{2}} [B_z\cos\Lambda-B_x\sin\Lambda]^2 \cdot \\
& \cdot \exp\left\{-\frac{|B_z\cos\Lambda-B_x\sin\Lambda|}{\beta_\parallel}\right\} \cdot \\
& \cdot \exp\left\{-\frac{|[(B_x\cos\Lambda+B_z\sin\Lambda)^2+B_y^2]^{\frac{1}{2}}|(1+\zeta\sin\Lambda)}{\beta_\perp}\right\} \;.
\end{split}
\label{equation:othercartesian}
\end{eqnarray}

This theoretical distribution is normalized to one as it verifies Equation \ref{equation:norma}. As in the two previous sections, 
we now perform a rotation of angle $\Theta$ around the $\ey$-axis to express Equation~\ref{equation:othercartesian} in the 
observer's reference frame, and then we transform into spherical coordinates in the observer's reference frame. This is done by 
applying Equations~\ref{equation:rotation} and \ref{equation:spherical}. A close inspection (e.g make for instance $\Lambda=0$) of 
Eq.~\ref{equation:othercartesian} shows that again the result does not depend on the choice of rotation axis in the $XY$-plane.
Although trivial to obtain, the resulting expression for the distribution function in spherical coordinates in the observer's 
reference frame is too long to be written here (we provide it for a simplified case below). It is noteworthy to mention, however, 
that even though Eq.~\ref{equation:othercartesian} depends on $\Lambda$, once this equation is expressed in the observer's reference 
frame, it depends both on $\Lambda$ and $\Theta$: $\mathcal{P}_1(\ve{B}; \Lambda, \Theta)$.\\

\begin{figure*}
\begin{center}
\begin{tabular}{cc}
\includegraphics[width=9cm]{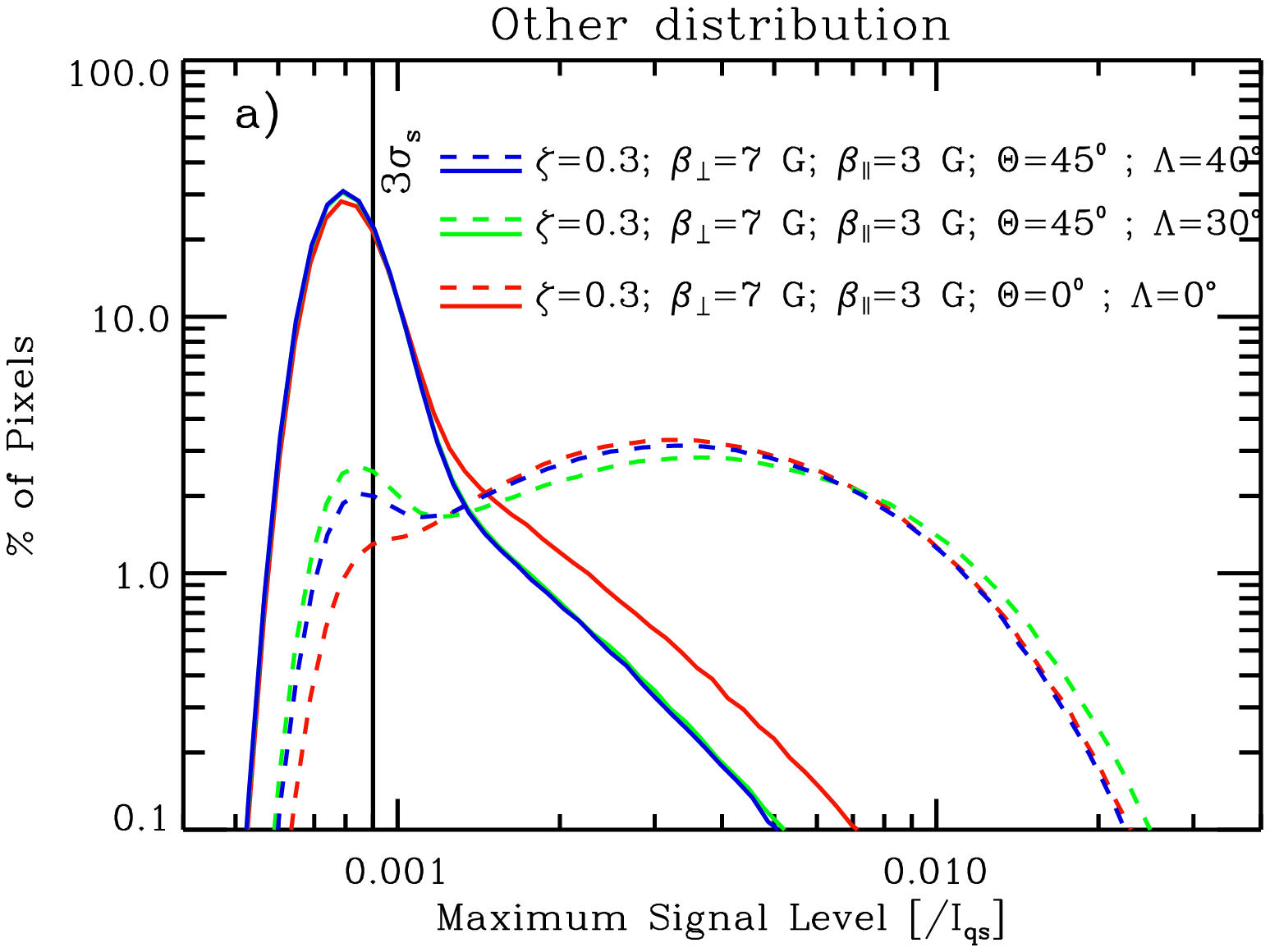} &
\includegraphics[width=9cm]{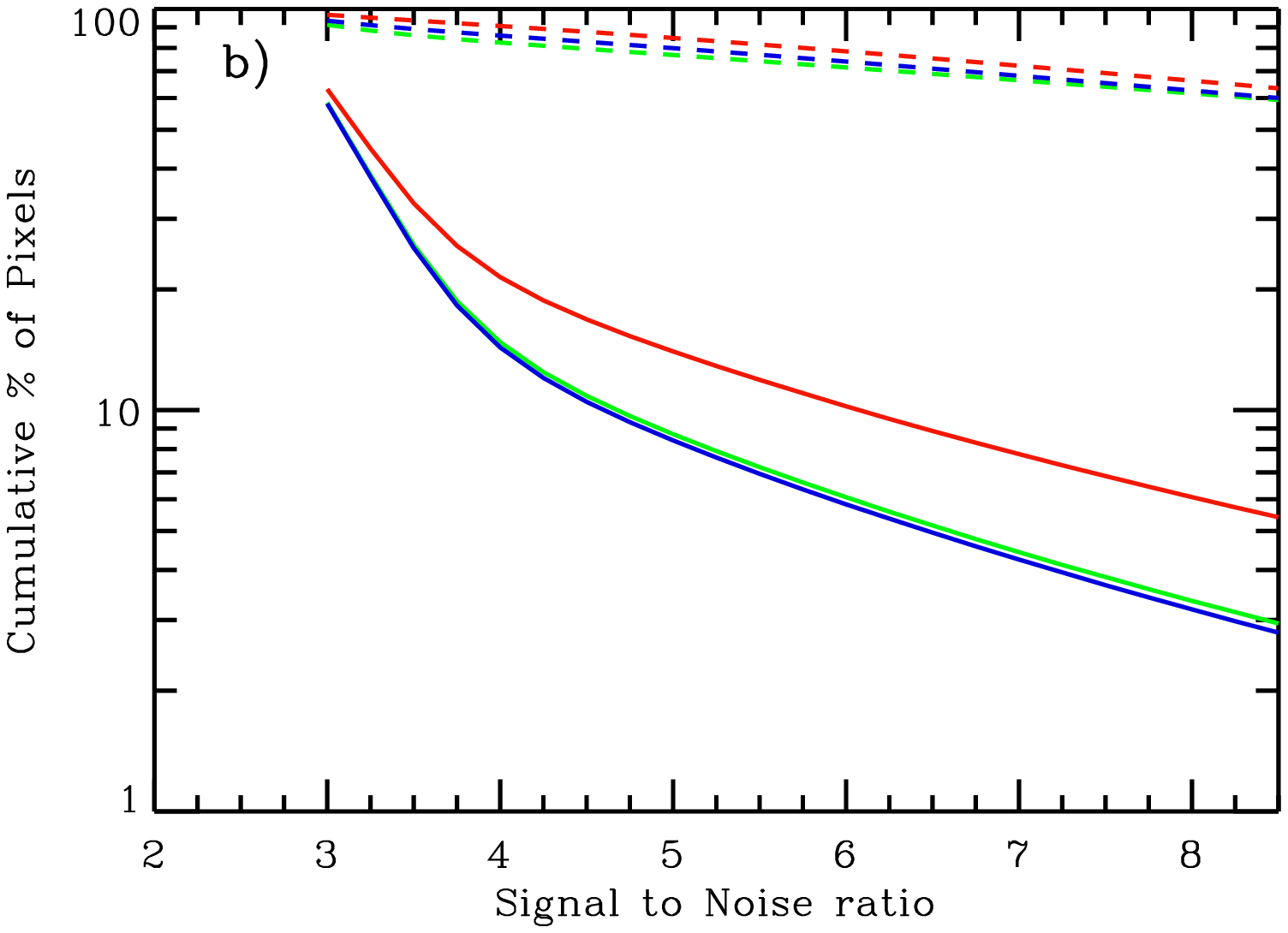} \\
\includegraphics[width=9cm]{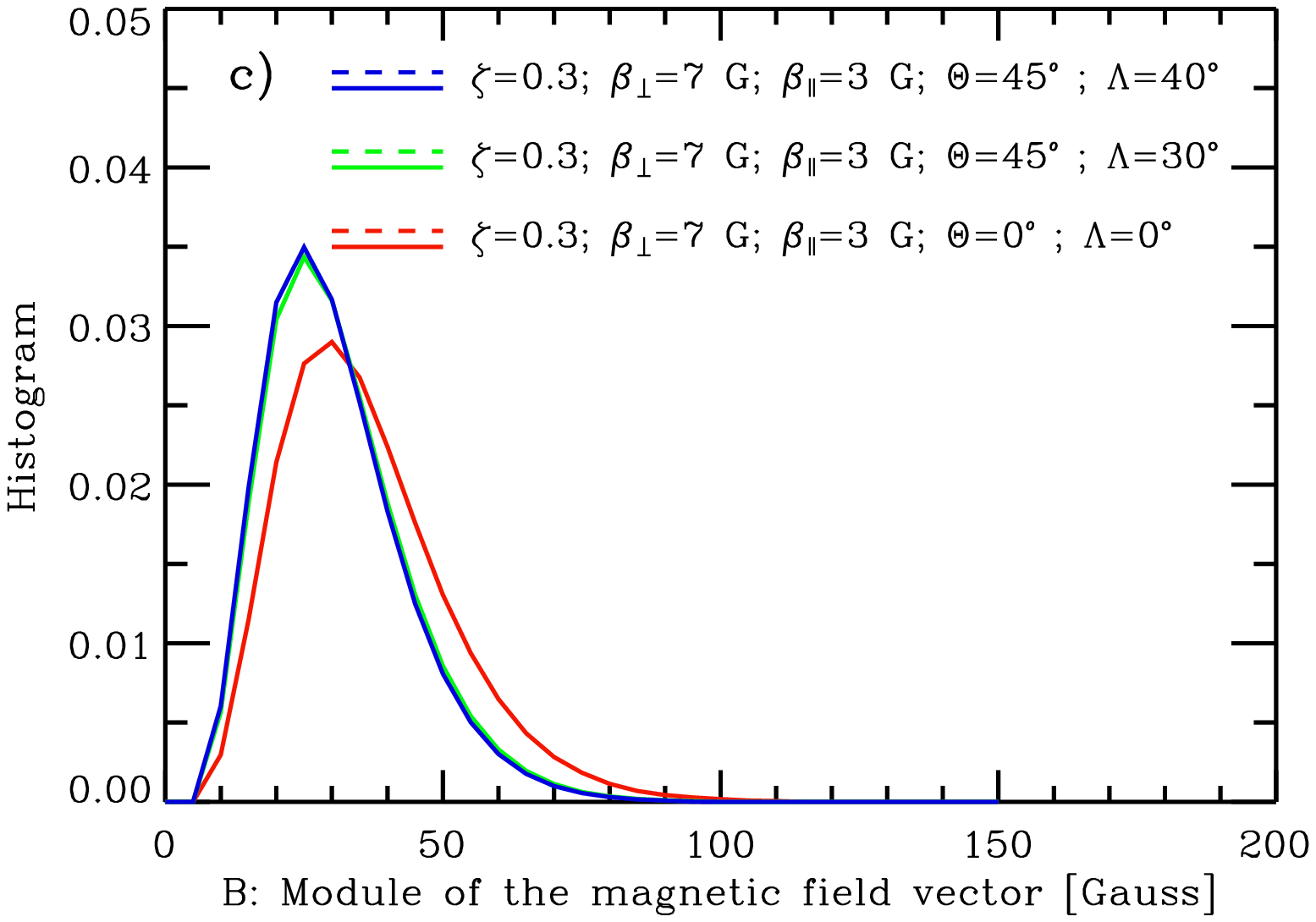} &
\includegraphics[width=9cm]{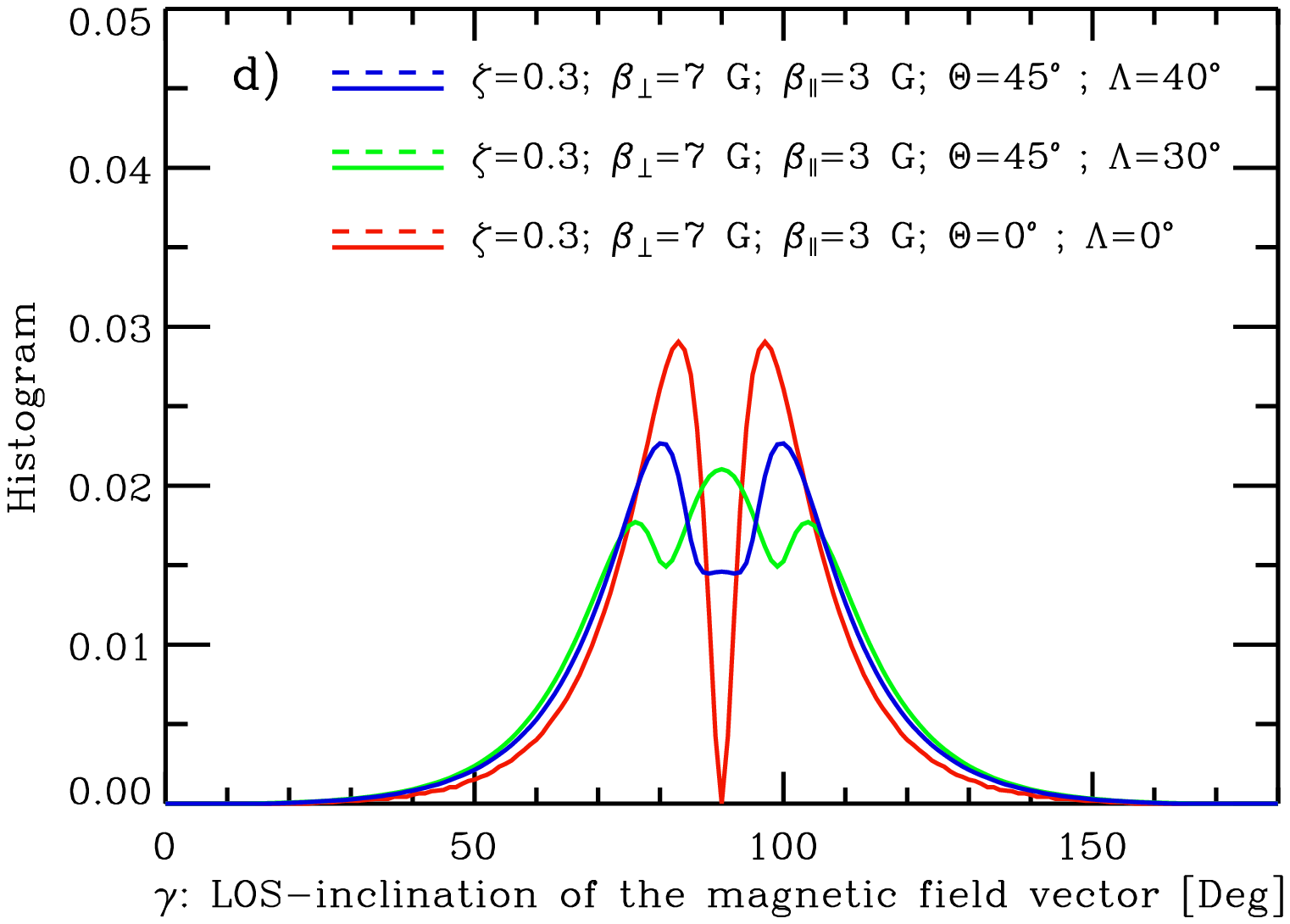}
\end{tabular}
\caption{Same as Figure~\ref{figure:isotropic} but for the theoretical distribution given by Eq.~\ref{equation:othercartesian}. The parameters
employed are $\beta_\parallel = 3$ G and $\beta_\perp=7$ G. Red lines correspond to $\Lambda=0\deg$ (disk center), while green and blue lines 
correspond to $\Lambda=30\deg$ and $\Lambda=40\deg$, respectively. Solid and dashed lines indicate the circular (Stokes $V$) and linear (Stokes $Q$ and $U$) 
polarization.}
\label{figure:other}
\end{center}
\end{figure*}

The theoretical distributions functions of $B$ and $\gamma$ resulting from Equation~\ref{equation:othercartesian} 
are shown in Figures~\ref{figure:other}c and \ref{figure:other}d, respectively. Here we considered
the following values: $\zeta=0.3$, $\beta_\parallel=3$, and $\beta_\perp=7$. Unlike Sections~\ref{subsection:iso} 
and \ref{subsection:triple}, $\beta_\parallel$ and $\beta_\perp$ do not correspond to the mean values of 
$B_\parallel$ and $B_\perp$. For this reason we refer to them as $\beta_\parallel$ and $\beta_\perp$ instead of 
$B_{\parallel 0}$ and $B_{\perp 0}$. In this new distribution, the module of the magnetic field vector $B$ is different 
at different latitudes, with a slightly higher mean value at disk center (red line; $\Lambda=0\deg$) 
than closer to the poles (green and blue lines; $\Lambda=30-40\deg$). This is a direct consequence of prescribing 
a distribution of the magnetic field vector in the local reference frame that depends on $\Lambda$.
Otherwise, as it occurred in Sects.~\ref{subsection:iso} and \ref{subsection:triple}, the distribution of $B$
would be the same, since the module of the magnetic field vector is invariant with respect to rotations.\\

We now use the distribution given by Equation~\ref{equation:othercartesian} to solve the radiative transfer equation 
and produce theoretical histograms of the Stokes profiles. These are displayed in Figures.~\ref{figure:other}a and 
\ref{figure:other}b. The shape of the curves for the circular (dashed) and linear (solid) polarization in Figure~\ref{figure:other}a 
are very similar to the observed ones (Fig.~\ref{figure:stokhistogram}a). Of particular interest is the fact that, 
in agreement with the observations, the histograms of the circular polarization are very similar in the three considered 
positions on the solar disk, and feature a peak at signals $\approx 3-4\times 10^{-3}$. It is also important to mention 
that the linear polarization decreases from $\Lambda=0\deg$ towards $\Lambda=30-40\deg$, which also agrees with 
the observations.\\

To understand how Equation~\ref{equation:othercartesian} can fit the observed histograms of the Stokes profiles at 
different latitudes, it is convenient to re-write this theoretical distribution function assuming that the observer's 
line-of-sight is along the zero meridian. In this case the rotation angle $\Theta$ is equal to $\Lambda$, thereby 
simplifying the probability distribution function from Eq.~\ref{equation:othercartesian} into (in spherical coordinates 
in the observer's reference frame)\\

\begin{eqnarray}
\begin{split}
\mathcal{P}_1 & (B,\gamma,\phi)\df B \df\gamma\df\phi = \frac{B^7\sin^4\gamma\cos^2\gamma}{192 \pi \beta_\parallel^3 \beta_\perp^5}
(1+\zeta\sin\Lambda)^5 \cdot \\ & \cdot \exp{\left\{-\frac{|B\sin\gamma|(1+\zeta\sin\Lambda)}{\beta_\perp}\right\}} \cdot \\ 
& \cdot \exp{\left\{-\frac{|B\cos\gamma}{\beta_\parallel}\right\}}  \df B \df \gamma \df \phi\;.
\end{split}
\label{equation:othersphe}
\end{eqnarray}

In this Equation~\ref{equation:othersphe}, the exponential term that refers to $B\cos\gamma$ (i.e. component of the magnetic field 
vector that is parallel to the observer's line-of-sight) does not depend on $\Lambda$ and therefore the amount of circular polarization 
generated by this distribution will not change with latitude (see dashed lines in Figs.~\ref{figure:other}a and \ref{figure:other}b). 
Moreover, the exponential term that contains $B\sin\gamma$ (i.e. component of the magnetic field vector that is perpendicular
to the observer's line-of-sight) decreases as $\Lambda$ increases and thus, this distribution function produces less linear polarization at higher 
latitudes (see solid lines in Figs.~\ref{figure:other}a and \ref{figure:other}b).\\

Despite all these similarities between the theoretical and observed distribution of Stokes profiles there are still some significant
differences, namely: {\bf a)} there is a clear deficit of pixels with sufficient amount
of linear polarization in the range of signals $\approx 2-4\times10^{-3}$ (Fig.~\ref{figure:other}a) or $S/R > 5$ (Fig.~\ref{figure:other}b)
at all latitudes; and {\bf b)} the theoretical histograms show a peak in the circular polarization at the $3\sigma$-level at high latitudes
(dashed-blue and dashed-green curves in Fig.~\ref{figure:other}a), while the observed histograms show no peak in the circular polarization at 
the $3\sigma$-level at any disk position (dashed curves in Fig.~\ref{figure:stokhistogram}a). Overall, however, the fit between the observed and 
the theoretical histograms of the Stokes profiles is clearly better for this distribution (Eq.~\ref{equation:othercartesian}) than for the 
case of an isotropic distribution (Eq.~\ref{equation:isotropic}) or a triple Gaussian (Eq.~\ref{equation:triplesphe}).\\

\section{Assumptions and limitations}
\label{section:limitations}

In the previous subsections we have focused on the effect that the probability distribution function of the magnetic field 
vector $\mathcal{P}_1(\ve{B},\alpha)$ has on the observed histograms of the Stokes profiles at different positions on the 
solar disk. In our theoretical analysis a number of simplifying assumptions were made to keep the problem tractable. Although 
they have already been pointed out in Section~\ref{section:pdftheory}, we summarize them here to briefly discuss their implications.\\

\begin{itemize}

\item We have assumed that the thermodynamic and magnetic parameters are statistically independent of each other. This allowed us 
to write the total probability distribution function in Eq.~\ref{equation:pdftot} as the product of two distinct probability 
distribution functions. However, as dictated by the Lorentz-force term in the momentum equation in magnetohydrodynamics, the 
magnetic field affects the thermodynamic structure of the solar atmosphere. It is therefore clear that this assumption does 
not fully hold in the solar atmosphere. For instance, if we take the commonly accepted picture of intergranular lanes harboring 
more vertical and stronger magnetic fields than the granular cells, and we consider that intergranular cells have a smoother
variation of the temperature with optical depth (see i.e. Fig.~3 in Borrero \& Bellot Rubio 2002), we could then have postulated 
a correlation between the magnetic field vector $\ve{B}$ and the gradient of the source function with optical depth $S_1$, which 
is contained in $\ve{T}$ (Eq.~\ref{equation:x}). Indeed, the higher $S_1$ is, the stronger will be the polarization profiles 
$Q$, $U$, and $V$ (see Eq.~9.45 in del Toro Iniesta 2003). These correlations could potentially cause the observed histograms 
of the Stokes profiles (Figure~\ref{figure:stokhistogram}) to vary with the heliocentric angle, even if the underlying distribution 
of the magnetic field vector does not depend on $\Theta$. Therefore it is important to investigate what an effect they have before 
conclusively proving that the distribution of the magnetic field vector is not isotropic (Sect.~\ref{subsection:iso}), or that the 
differences in the observed histograms of the Stokes profiles are not due to the viewing angle (Sect.~\ref{subsection:triple}).
Unfortunately, the aforementioned correlations are not known for the solar internetwork simply because it is not clear how magnetic 
fields are distributed here. In the future we will explore this question by employing 3D numerical simulations of the solar 
atmosphere, because they provide correlations between $\ve{B}$ and $\ve{T}$ that are compatible with the MHD equations.\\

\item We have also assumed that the probability distribution function of the thermodynamic and kinematic parameters, 
$\mathcal{P}_2(\ve{T},V_{\rm los})$, does not depend on the position on the solar disk. Kinematic parameters (i.e. line-of-sight 
velocity $V_{\rm los}$) do not influence our study, since they have no effect on the amplitude of the Stokes profiles in 
Fig.~\ref{figure:stokhistogram}. The same can be argued about other thermodynamic parameters in $\ve{T}$, such as the source 
function at the observer's $S_0$ (affects only Stokes $I$), and damping parameter $a$ (affects mostly the line width but 
not its amplitude). By far, the most important thermodynamic parameters affecting the amplitude of the Stokes profiles under 
the Milne-Eddington approximation are the gradient of the source function with optical depth $S_1$ and the continuum-to-line-center 
absorption coefficient $\eta_0$. In a 1D atmosphere both these parameters are known to decrease as $\Theta$ increases, because the 
line-of-sight samples a thinner vertical-portion of the atmosphere. However, since the dependence of the polarization profiles with 
$S_1$ and $\eta_0$ are identical (see Eqs. 8.14, 8.15 and 9.44 in del Toro Iniesta 2003), one would expect the same drop with
increasing heliocentric angle angle in the amplitude of the circular polarization profiles (Stokes $V$) and linear polarization 
profiles (Stokes $Q$ and $U$). However, Figure~\ref{figure:stokhistogram} shows that the linear and circular polarization profiles 
(solid-color and dashed-color lines) behave differently, and therefore we can rule out the variations of $S_1$ and/or 
$\eta_0$ with $\Theta$ as being responsible for the observed histograms in the Stokes profiles. Of course, this would change in a 
3D atmosphere, where the line-of-sight pierces through different inhomogeneous atmospheric layers, thereby opening the door for the 
possibility of $S_1$ and or $\eta_0$ to affect the linear and circular polarization profiles differently\footnote{Outside the 
Milne-Eddington approximation, a similar argument could be made in terms of the stratification with optical depth of the 
temperature in the solar photosphere $T(\tau_c)$, instead of $S_1$ and $\eta_0$.}.\\

\item Adopting a Milne-Eddington atmosphere also implies that we are assuming that the magnetic field vector $\ve{B}$ does not 
vary with optical depth $\tau_c$ in the photosphere. This can have important consequences, since at larger heliocentric angles 
the spectral line samples higher atmospheric layers than at disk center, where the probability distribution function of the 
magnetic field vector can be different. Employing the widely used 1D HOLMU model (Holweger \& M\"uller 1974) we calculated that 
the continuum level $\tau_c=1$ rises by approximately $20$ Km from disk center $\Theta=0\deg$ (map A; Sect.~\ref{subsection:mapa}) 
to $\Theta=45\deg$ (maps B and C; Sects.~\ref{subsection:mapb} and \ref{subsection:mapc}). Since this vertical shift of the continuum 
level is rather small, we could argue that the histograms of the Stokes profiles in Fig.~\ref{figure:stokhistogram} are not affected 
by this effect. However, the value of 20 km should be considered only as a lower limit since a 1D model does not take into account 
the horizontal inhomogeneities present in the solar atmosphere. To properly account for this effect, more sophisticated 3D models should 
be employed.\\

\item Finally, we have considered $\alpha=1$ in our analysis (see Sect.~\ref{section:pdftheory}). This is equivalent to considering 
that, at the resolution of the Hinode/SP instrument (0.32"; Sect.~\ref{section:observations}), the magnetic structures are spatially 
resolved. This is, of course, highly unlikely, and therefore it would be important to drop this assumption in the future. Its importance 
can only be quantified with additional assumptions about the scale-distribution of the magnetic structures in the solar photosphere. 
This topic is, in itself, as controversial as the distribution of the magnetic field strength and inclination, which is the reason
why we have refrained from addressing it here. Although employing 3D MHD simulations would certainly help to drop the $\alpha=1$ assumption, 
we are cautious about it since it is not clear whether these simulations are reliable at the smallest physical scales (S\'anchez Almeida 2006).\\

\end{itemize}

\section{Discussion and conclusions}
\label{section:conclu}

The histograms of the observed Stokes profiles at different positions on the solar disk (Fig.~\ref{figure:stokhistogram})
are clearly different from each other. One possible interpretation for this is that the distribution of the magnetic 
field vector in the solar internetwork is not isotropic. We explored this possibility in Section~\ref{subsection:iso},
where we employed an isotropic probability distribution of the magnetic field vector. This distribution yielded, as expected, 
the same distribution of Stokes profiles at all positions on the solar disk (Fig.~\ref{figure:isotropic}).
Mart{\'\i}nez Gonz\'alez et al. (2008) have also presented similar histograms but employing the Stokes
profiles from the Fe I line pair at 1.56 $\mu$m (observed with the TIP2 instrument; Mart{\'\i}nez Pillet et al. 1999).
Their histograms (see their Figure 2) showed no clear variation with the heliocentric angle, which lead them to 
conclude that the distribution  of the magnetic field vector in the quiet Sun was isotropic. Interestingly, these authors 
also mentioned after a more detailed analysis that there could indeed be a dependence of the histograms with the heliocentric angle
(as indeed we find here).\\

In addition to Mart{\'\i}nez Gonz\'alez et al. (2008), a number of works have also argued in favor of an 
isotropic distribution of magnetic fields in the internetwork. In particular, Asensio Ramos (2009) 
and Stenflo (2010), employing two different approaches, both concluded that for very weak magnetic fields ($B \to 0$) the distribution 
becomes isotropic. With our present data we cannot argue against or in favor of this interpretation. The main reason for this is that,
as discussed in Section~\ref{section:clvobs}, any distribution for the magnetic field vector where $B_\perp$ has a peak below 40-70 
will produce linear polarization profiles that are dominated by noise ($3\sigma$-level or $S/R = 3$). Therefore our current approach
(described in Section~\ref{section:pdftheory}) cannot be employed to discern the underlying distribution of the magnetic field vector
from these profiles dominated by noise. However, it can be employed to establish that the number of pixels that would follow 
this hypothetically isotropic distribution cannot be much larger than 30 \% of the pixels in the internetwork, since this is the amount of pixels 
that show a peak at the $3\sigma$-level in the polarization profiles (see Fig.~\ref{figure:stokhistogram}a). For signals above $> 2\times 10^{-3}$, 
the histograms of the Stokes profiles deviate significantly from the ones predicted by an isotropic distribution, and thus we can establish
that here the distribution of the magnetic field vector cannot be isotropic.\\

We can use a different argument to further clarify the previous point. Our theoretical distributions in Section~\ref{section:pdftheory}
apply to all possible values of the module of the magnetic field vector. However, we could have employed distributions pieced together 
in the following form:

\begin{eqnarray}
\mathcal{P}_1(\ve{B})\df \ve{B} = \begin{cases} \mathcal{P}_a(\ve{B})\df\ve{B}, & \textrm{if}\; B<B^{*} \\ 
\mathcal{P}_b(\ve{B})\df \ve{B} , & \textrm{if}\; B>B^{*} \;, \end{cases}
\label{equation:piecepdf}
\end{eqnarray}

\noindent where $\mathcal{P}_a$ could hypothetically correspond to an isotropic distribution for weak fields: $B<B^{*}$. This would explain 
the $3\sigma$-peak in the linear polarization in Figure~\ref{figure:stokhistogram}a (dashed lines). In addition, $\mathcal{P}_b$ could be 
a distribution, valid for larger fields $B > B^{*}$, that would fit the tails of the histogram. The distribution given by Equation~\ref{equation:piecepdf}
does not need to be discontinuous because it could be prescribed such that $\mathcal{P}_a(B^{*}) = \mathcal{P}_b(B^{*})$.\\

In Section~\ref{subsection:triple} we employed a triple Gaussian (one for each component of the magnetic field vector) distribution
function and found that, under this assumption and at disk center, the best fit to the observed histograms of the Stokes profiles 
is produced by a distribution in which the mean value of the magnetic field vector component that is parallel to the solar surface 
is lower than the mean value of the magnetic field vector component that is perpendicular to the solar surface: 
$B_{z0} < \sqrt{B_{\rm x0}^2+B_{\rm y0}^2}$. This yields a distribution function where the magnetic field vector is highly inclined, 
in agreement with previous findings from Orozco et al. (2007a, 2007b) and Lites et al. (2007, 2008). However, this distribution does 
not fit well the histograms of the Stokes profiles at other positions on the solar disk. In fact, in that section we found that 
it is not possible to fit the observed histograms for the Stokes profiles at different heliocentric angles employing a theoretical 
distribution function for the magnetic field vector prescribed in the local reference frame that only changes due to the viewing 
angle $\Theta$. The reason for this is that, for an underlying distribution where the magnetic field vector is mostly horizontal
at $\Theta=0\deg$ (disk center), the amount of linear polarization slightly decreases when $\Theta$ increases, while the amount of 
circular polarization would significantly increase as $\Theta$ increases. However, the observed histograms of the Stokes profiles (Fig.~\ref{figure:stokhistogram}) show that, although the amount of linear polarization decreases when $\Theta$ increases, the circular polarization 
does not particularly increase (see also discussion in Lites et al. 2008). This cannot be explained in terms of a simple rotation of the 
viewing angle $\Theta$, and therefore we interpreted this fact, in Section~\ref{subsection:other}, as proof that the underlying (i.e. in 
the local reference frame) distribution of the magnetic field vector must depend on the position on the solar disk.\\

Under the assumption that the distribution of the underlying magnetic field vector depends on the latitude $\Lambda$ 
(see Sect.~\ref{subsection:other}), we were able to find a theoretical distribution of the magnetic field vector 
(Eq.~\ref{equation:othercartesian}) that fits quite well the observed histograms of the Stokes profiles at different 
positions on the solar disk (Figure~\ref{figure:other}). Among other properties, this distribution features a magnetic field 
whose mean value decreases toward the poles. We note here that this does not mean that this is the real distribution 
for the magnetic field vector present in the quiet Sun. One reason for  this is that the fit is far from perfect (see 
discrepancies mentioned in Sect.~\ref{subsection:other}), but most importantly, that we do not know whether this solution 
is unique because there can be other theoretical distributions that fit the observed Stokes profiles equally well, or even better. 
More work is indeed needed to confirm or rule out Eq.~\ref{equation:othercartesian} as the real distribution of 
the magnetic field vector present in the Sun. In particular, a better fit to the observed histograms of the Stokes profiles 
is desirable. In addition, it is important to have maps at more latitudes to further constrain the possible distribution functions.\\

Moreover, it is important to bear in mind that the conclusions above are not necessarily the only possible interpretations, 
because postulating a probability distribution function of the thermodynamic and kinematic parameters, $\mathcal{P}_2(\ve{T},V_{\rm los})$, 
that varies with the heliocentric angle $\Theta$, or postulating a correlation between the thermodynamic ($\ve{T}$) and 
magnetic ($\ve{B}$) parameters might also help explain the observed differences between the histograms of the Stokes profiles 
at different positions on the solar disk. Another effect that has not been accounted for is that magnetic field vector can vary with
optical depth in the solar photosphere. Since the Stokes profiles sample increasingly higher atmospheric layers as the 
heliocentric angle increases, the distribution of the magnetic field vector can be different for different values of $\Theta$, 
even if the probability distribution of the magnetic field vector is the same at all positions on the solar disk at a fixed 
geometrical depth. All these effects could be properly accounted for by means of 3D MHD simulations of the solar photosphere 
(Sch\"ussler \& V\"ogler 2008; Steiner et al. 2008, 2009; Danilovic et al. 2010).\\

In the future we expect to employ such simulations to either rule out or confirm our results in this paper. Consequently, 
our conclusions at this point should be regarded as preliminary only. Instead, the main purpose in this paper is to illustrate 
the methodology detailed in Sections~\ref{section:clvobs} and \ref{section:pdftheory} to study the distribution of the magnetic 
field vector in the quiet Sun, by directly inverting the histograms of the Stokes profiles in entire maps instead of 
inverting the Stokes profiles at each spatial position in a given map. Our method has great potential to investigate several 
aspects of the photospheric magnetism  in the solar internetwork. For instance, it can be used, as in Sect.~\ref{subsection:other}, 
to confirm whether the mean value of the distribution of the magnetic field vector changes from disk center toward the poles 
(cf. Zwang 1987; Ito et al. 2010). This will have important consequences for theoretical models that explain
 the torsional oscillations in the butterfly diagram in terms of a geostrophic flow model (Spruit 2003), which requires a 
significant amount of magnetic flux at high latitudes at the beginning of the sunspot cycle. In addition, and although in 
this work we have restricted ourselves to variations in latitude ($\Lambda$), additional observations from disk center 
toward the solar limbs could be employed to investigate whether the properties of the 
magnetic field in the internetwork change also in longitude. This is already predicted by non-axisymmetric dynamo models (Moss 1991, Moss et al. 1999, 
Bigazzi \& Ruzmaikin 2004, Charbonneau 2005) and can provide important clues about the strength of the differential rotation (R\"udiger \& Elstner 1994; 
Zhang et al. 2003).\\

\begin{acknowledgements}
We would like to thank Luis Bellot Rubio, Mariam Mart{\'\i}nez Gonz\'alez and Oskar Steiner for fruitful 
discussions in the subject. Many thanks also to an anonymous referee, who pointed out an error in the probability
distribution function of Section~\ref{subsection:triple} in an early version of this manuscript. This work analyzes 
data from the Hinode spacecraft. Hinode is a Japanese mission developed and launched by ISAS/JAXA, 
collaborating with NAOJ as a domestic partner, NASA and STFC (UK) as 
international partners. Scientific operation of the Hinode mission is conducted by the Hinode science team 
organized at ISAS/JAXA. This team mainly consists of scientists from institutes in the partner countries. 
Support for the post-launch operation is provided by JAXA and NAOJ (Japan), STFC (U.K.), NASA, ESA, and NSC (Norway). 
This work has also made use of the NASA ADS database.
\end{acknowledgements}

\end{document}